\documentclass[aps,prd]{revtex4}
\usepackage{epsfig,epsf}
\usepackage{amsmath}
\usepackage{amsthm}
\usepackage{amsfonts}
\usepackage{amssymb}
\usepackage{dsfont}
\usepackage{multirow}
\usepackage{appendix}
\usepackage{slashed}
\usepackage[active]{srcltx}
\usepackage{psfrag}
\usepackage{subfigure}
\usepackage[colorlinks,citecolor=blue,urlcolor=red,linkcolor=red]{hyperref}
\usepackage[colorlinks,citecolor=blue,urlcolor=red,linkcolor=red]{hyperref}
\begin{document}

\title{{\Large{\bf Semileptonic $D_{(s)} \to A \ell^+ \nu$ and nonleptonic $D\to K_1
(1270,1400) \pi$ decays in LCSR}}}

\author{\small
S. Momeni\footnote {e-mail: samira.momeni@ph.iut.ac.ir}, R.
Khosravi\footnote {e-mail: rezakhosravi @ cc.iut.ac.ir}}

\affiliation{Department of Physics, Isfahan University of
Technology, Isfahan 84156-83111, Iran }

\begin{abstract}
In this work, the  transition form factors are calculated for the
semileptonic $D_{(s)} \to A \ell^+ \nu$ where $A=a_{1}, b_{1},
K_{1}(1270,1400)$, i.e., $D^{+} \to a^{0}_{1} (b^{0}_{1}, K^{0}_{1})
\ell^+ \nu$, $D^{0} \to a^{-}_{1} (b^{-}_{1}) \ell^+ \nu$,  and
$D^{+}_{s}\to K^{0}_{1} \ell^+ \nu$ in the frame work of the
light-cone QCD sum rules (LCSR) approach up to the twist-3
distribution amplitudes (DAs). Since the masses of these axial
vector mesons are comparable to the charm quark mass, we keep out in
our calculations all terms including ${m_{A}}/{m_c}$ in expansion of
two--parton DAs. Branching ratio values are estimated for the
semileptonic $D_{(s)} \to A \ell \nu$ and nonleptonic $D\to K_1
(1270,1400) \pi$ decays. A comparison is also made between our
results and predictions of other methods and the existing
experimental values.
\end{abstract}

\pacs{11.55.Hx, 13.20.-v, 14.40.Lb}

\maketitle

\section{Introduction}
Exclusive semileptonic decays of $B$ and $D$ mesons are very useful
to determine the weak interaction couplings of quarks within the
standard model (SM) because of their relative abundance and the
simplicity of their study comparative to nonleptonic decays. In
connection with the charmed hadrons, there are several features that
make a difference between them and other hadrons. In the following,
some of them are mentioned \cite{Artuso}:

1) The  mass of the charmed hadrons is about $2~ \rm GeV$.  In these
region, the nonperturbative hadronic physics is operative. However,
theoretical methods developed for heavy quarks can in principle
still be applied, albeit with larger uncertainties.

2) Charmed hadron data can be used to probe the Yukawa sector of the
SM by the lattice QCD simulations.

3) In many cases, charm transitions provide almost background-free
low-energy signals of new physics (NP).

4) Manifestation for charm quark existence came from low energy kaon
oscillation experiments. Similar to this, one can hope that
oscillations of charmed hadrons can provide hints of what is
happening at the TeV scale.

The meson $D_{(s)}$, which contains one heavy charm quark $c$ and
one light quark, is placed in the heavy mesons category. The heavy
charmed meson can decay into the axial vector mesons by emitting a
pair of leptons $\ell \nu$ through weak interaction. In quark level,
this process is induced by the semileptonic decay of charm quark $c
\to q \ell \nu$, where $q=d, s$. This light quark $d$ or $s$ is
bound with the initial light quark in the charmed meson by strong
interaction to form an axial vector meson. It should be noted that
another category of charmed meson decays can be fulfilled by the
flavor changing neutral currents (FCNC) at tree level in SM via the
$c \to u \ell^+ \ell^-$ transition such as $D \to \pi \ell^+
\ell^-$, $D \to \rho \ell^+ \ell^-$, $D \to \pi \,\gamma$ and $D \to
\rho\, \gamma $ decays (for more detailed, see \cite{Feldmann}).
Analyzing the semileptonic decays of the charmed $D_{(s)}$ meson is
important for determination of the Cabibbo-Kabayashi-Maskawa (CKM)
matrix elements, checking the standard model and also calculation of
the leptonic decay constants of the initial and final meson states.

Nonperturbative effects of semileptonic decays can be parameterized
by transition form factors. Considering the transition form factors
for the semileptonic decays of mesons has two-fold importance:

$\bullet$ A number of the physical observables  such as decay withe
and branching ratio, in addition some parameters of the SM can be
investigated using these form factors.

$\bullet$ The factorization of amplitudes in the nonleptonic
two-body decays can be fulfilled in terms of the transition form
factors.

The form factors are calculated by various methods. Each method is
more powerful than other methods in a certain region of the
transferred momentum square $q^2$. In the region $q^2\rightarrow 0$
where the momentum of the outgoing meson is high, the large energy
effective theory (LEET) can be used for determining of the form
factors. In the region of large momentum transfer ($q^2 \rightarrow
q^2_{ max}$), the lattice QCD (LQCD) can be used. In $q^2 = q^2_{
max}$, the Form factors can be calculated within the heavy-light
chiral perturbation theory $(HL_{\chi}PT)$, that is based on the
heavy quark effective theory (HQET). Calculations within
$HL_{\chi}PT$ can also be completed by calculations in the frame
work of the heavy-light chiral quark model ($HL_{\chi}QM$).
Corrections via the heavy quark symmetry are of the order
$\mathcal{O}(1/m_c)$, which will be larger in the $D$ sector than in
the $B$ sector \cite{Palmer}.

A particularly approach to heavy-to-light transitions is offered by
QCD sum rules on the light-cone. The LCSR approach combine operator
product expansion (OPE) on the light-cone with QCD sum rule
techniques in the region that $q^2$ is near zero. In this approach,
the nonperturbative hadronic matrix elements are described by the
light-cone distribution amplitudes (LCDAs) of increasing twist
instead of the vacuum condensates (for more details, see Refs.
\cite{Chernyak,Kolesnichenko,Filyanov,Zhitnitsky,Filyanov2}).

The form factors of the semileptonic decays of charmed meson
$D_{(s)}$ to scalar, pseudoscalar or vector mesons have been
estimated by various approaches. The form factors of the
semileptonic decays $D^+ \to (D^0, \rho^0, \omega, \eta,
\eta')\ell^+ \nu$ and $D^+_s \to (D^0, \phi, K^0, K^{*0}, \eta,
\eta') \ell^+ \nu $ have been computed in the framework of the
covariant confined quark model (CCQM) \cite{Soni1,Soni2}. Both the
vector and scalar form factors of $D\to K \ell \nu$ decay have been
determined from the experimental measurements \cite{Zhang}. In Ref.
\cite{Khodjamirian, BallD,Fu}, the $D\to \pi(K, \rho)\,\ell\, \nu$
decays have been studied by the LCSR approach. The semileptonic
processes $D\to \pi, \rho, K$ and $K^{*}$ have been investigated by
the HQET in Ref. \cite{WangD}, while the form factors of the $D\to
\pi (K, K^{*})\ell\,\nu$ transitions have been evaluated by the LQCD
method in Ref. \cite{Abada, Aubin, Bernard}. The semileptonic decays
$D_{(s)} \to f_0 (K_0^*)\,\ell\, \nu$, $D_{(s)} \to \pi (K) \,\ell\,
\nu$, and $D_{(s)} \to K^* (\rho, \phi) \,\ell\, \nu$  have been
studied in the framework of the three-point QCD sum rules (3PSR)
\cite{Ignacio,Aliev, Ball2,Ball3,Ovchinnikov,Dong,Mao}. For the
axial vector meson, as the final state in $D$ meson decays, the
$D_{q}\to K_{1}\,\ell\,\nu\,(q=u, d, s)$ and
$D\to\,a_{1},f_{1}(1285), f_{1}(1420)$ transitions have been
analyzed by the 3PSR approach \cite{Khosravi,Zuo}.

The main purpose of this paper is the form factor investigation for
the semileptonic decays of $D_{(s)}$ meson to the axial vector
mesons such as: $D^{0} \to a^{-}_{1} (b^{-}_{1}) \ell^+ \nu$, $D^{+}
\to a^{0}_{1} (b^{0}_{1}) \ell^+ \nu$, $D_{s}^{+} \to K^{0}_{1}
\ell^+ \nu$ as well as $D^{+}\to K^{0}_{1} \ell^+ \nu$ decays. The
first three cases of these decays are described by $c \to d~ \ell
\nu$ transition at quark level, while the latter is proceed by $c
\to s~ \ell \nu$ transition. We plan to calculate the form factors
of the aforementioned semileptonic decays up to the twist-3 DAs of
the axial vector mesons in the framework of the LCSR. It should be
noted that we keep out all terms including $m_{A}/m_c$ ($m_{A}$
stands for the axial vector mass) in the expansion of the
two-particle DAs of the axial vector mesons since their masses are
comparable to quark mass $m_c$. We compare our results for the
transition form factors of the semileptonic decays  with predictions
obtained from other methods. Using the computed  form factors, the
branching ratios of the nonleptonic $D^0 \to K^-_1(1270) \pi^+$,
$D^0 \to K^-_1(1400) \pi^+$, $D^+ \to K^0_1(1270)\pi^+$ and $D^+ \to
K^0_1(1400)\pi^+$ decays are considered. A comparison is made
between  our values for the branching ratios of the aforementioned
nonleptonic decays with results obtained form other approaches as
well as existing experimental values.

The physical states of $K_1(1270)$ and $K_1(1400)$ mesons are
considered as a mixture of two $|K_{1A}\rangle$ and $|K_{1B}\rangle$
states and can be parameterized in terms of a mixing angle
$\theta_K$, as follows \cite{Kwei2}:
\begin{eqnarray}\label{eq21}
|K_1(1270)\rangle &=&\sin\theta_K |K_{1A}\rangle +
\cos\theta_K |K_{1B}\rangle,\nonumber\\
|K_1(1400)\rangle &=&\cos\theta_K |K_{1A}\rangle - \sin\theta_K
|K_{1B}\rangle,
\end{eqnarray}
where $|K_{1A}\rangle$ and $|K_{1B}\rangle$ have different masses
and decay constants. Also, the mixing angle $\theta_K$  can be
determined by the experimental data. There are various approaches to
estimate the mixing angle. The result $35^\circ \leq |\theta_K| \leq
55^\circ$ was found in Ref. \cite{Burakovsky}, while two possible
solutions were obtained as $|\theta_K|\approx 33^\circ \vee
57^\circ$  in Ref. \cite{Suzuki} and as $|\theta_K|\approx 37^\circ
\vee 58^\circ$ in Ref. \cite{HYCheng}. A new window for the value of
$\theta_K$ is estimated from the results of $B\to K_1(1270)\gamma$
and $\tau \to ¨ K_1(1270)\nu_{\tau}$ data as \cite{Hatanaka2}
\begin{eqnarray}
\theta_K = {-(34\pm13)}^{\circ}.
\end{eqnarray}
Sofar this value is used in Refs.
\cite{Dag,Bayar,Yang1,Bashiry,Falahati,Hatanaka}. In this study, we
also use the result of $\theta_K = {-(34\pm13)}^{\circ}$.

The paper is organized as follows: In Sec. \ref{se.2}, by using the
LCSR method, the form factors for the semileptonic decays of $D$ to
the axial vector mesons are derived. In Sec. \ref{se.nu}, we present
our numerical analysis for the form factors and determine the
branching ratio values of the semileptonic and nonleptonic decays. A
comparison is also made between  our results and the predictions of
other methods in this section.

\section{Transition form factors in the LCSR}\label{se.2}

To calculate the form factors of the semileptonic transition of
$D^0$ to the axial vector meson $a_{1}^{-}$ $(D^0 \to a_{1}^{-}
\ell^{+} \nu)$ in the LCSR method, we consider the following
correlation function as
\begin{eqnarray}\label{eq24}
\Pi_\mu (p , p')&=&i \int d^4x\, e^{iqx} \langle a_{1}^{-} (p',
\varepsilon) | {\cal{T}} \{\bar{d}(x) \gamma_\mu (1-\gamma_5) c(x)\,
j_{D^0}^{\dag}(0) \}| 0 \rangle,
\end{eqnarray}
in this correlation function, $q=p-p'$, where $p$ and $p'$ are the
four-momentum of the initial and final meson states, respectively.
In addition, $j_{D^0}=i \bar{u} \gamma_5 c$ is known as the
interpolating current of $D^0$ meson. Current  $\bar{d} \gamma_\mu
(1-\gamma_5) c$ is the interaction current for semileptonic $D^0\to
a^{-}_{1}$ transition.

Following  the general idea of the LCSR, the correlation function in
Eq. (\ref{eq24}) should be calculated in two different languages: 1)
in terms of hadronic properties which we say the physical
representation, and 2) quark and gluon degrees of freedom which is
the theoretical side. Equating two sides and applying the Borel
transformation to suppress the contribution of the higher states and
continuum, we get sum rule expressions for the form factors. Let us
first consider the physical representation of the correlator
function.

To obtain the phenomenological or physical representation of the
correlation function, a complete set of intermediate states with the
same quantum numbers as the current $J_{D^0}$ is inserted between
two currents in Eq. (\ref{eq24}). Isolating the pole mass term of
the pseudoscalar $D^0$ meson and  applying Fourier transformation,
we get
\begin{eqnarray}\label{eq25}
\Pi_{\mu} (p',p)&=& \frac{\langle a_1^- (p',\varepsilon)|\bar
{d}\,\gamma_{\mu}(1-\gamma_5)\,c |D^0(p)\rangle \langle D^0
(p)|\bar{c}~ i \gamma_5\,u|0\rangle}{m^2_{D^0}-p^2}+ \mbox{higher
states and continuum}.
\end{eqnarray}
The matrix element $\langle a_1^- (p',\varepsilon) | \bar{d}
\gamma_\mu (1-\gamma_{5}) c | D^0 (p)\rangle$ is parameterized in
terms of the form factors as follows:
\begin{eqnarray}\label{eq27}
\langle a_1^{-} (p',\varepsilon) |\bar{d} \gamma_\mu
(1-\gamma_{5})c| D^0(p) \rangle &=& i  {2 A (q^2) \over m_{D^0}-
m_{a_1^{-}}} \epsilon_{\mu\nu\alpha\beta}\, \varepsilon^{*\nu}
p^\alpha p'^\beta - V_{1}(q^2)  \varepsilon_\mu^{*}(m_{D^0} -
m_{a_1^{-}}) \nonumber
\\&+&
\frac{V_{2}(q^2)}{m_{D^0}-m_{a_1^{-}}} (\varepsilon^{*}. q) (p+p')_\mu
+ \frac{( \varepsilon^{*}.q) 2 m_{a_1^{-}}}{q^2} q_\mu [V_{3}(q^2) -
V_{0}(q^2)],
\end{eqnarray}
where $m_{a_1^{-}}$ and $\varepsilon_\mu$ are the mass and the
four-polarization vector of the axial vector meson $a_1^{-}$,
respectively. In Eq. (\ref{eq27}), $A(q^2)$ and
$V_{i}(q^2)~~(i=0,...,3)$ are the transition form factors of the
$D^0\to a_1^{-} \ell^+ \nu$ decay. Form factor $V_{3}(0) $ can be
written as a linear combination of $V_{1}(q^2) $ and $V_{2}(q^2) $
as
\begin{eqnarray}\label{eq28}
V_{3}(q^2)=\frac{m_{D^0}-m_{a_1^{-}}}{2m_{a_1^{-}}}\, V_{1}(q^2)
-\frac{m_{D^0}+m_{a_1^{-}}}{2m_{a_1^{-}}}\, V_{2}(q^2),
\end{eqnarray}
with the condition $V_{0}(0)=V_{3}(0)$.

The second matrix element in Eq. (\ref{eq25}) is expressed in the
standard way as
\begin{eqnarray}\label{eq28d}
\langle D^0(p)|\bar{c}\,i \gamma_5\,u\,|0\rangle=\frac{f_{D^0}
m_{D^0}^{2} }{m_c+m_{d}},
\end{eqnarray}
where $f_{D}$ is the $D$ meson decay constant and $m_c(m_{d})$ is
the $c(d)$ quark mass. Using Eqs. (\ref{eq27}) and (\ref{eq28d}) in
Eq. (\ref{eq25}), the phenomenological part of the correlation
function is written in terms of the form factors and Lorentz
structures as
\begin{eqnarray}\label{eq29}
\Pi_\mu&=& -\frac{f_{D^0} m_{D^0}^2}{m_{c}+m_{d}} \frac{1}{p^2 -
m_{D^0}^2} \Bigg\{i \frac{2 A(q^2)}  {m_{D^0} - m_{a_1^{-}}}
\epsilon_{\mu\nu\alpha\beta} \varepsilon^{*\nu} p^\alpha p'^\beta
-V_1(q^2) \varepsilon_\mu^{*}(m_{D^0}-m_{a_1^{-}}) \nonumber \\
&+& \frac{V_2(q^2)}{ m_{D^0}-m_{a_1^{-}}}(\varepsilon^{*}. q)
{(p+p')}_\mu + \frac{(\varepsilon^{*}.q) 2 m_{a_1^{-}} }{q^2}
q_{\mu} [V_3(q^2)-V_0(q^2)] \Bigg\}+ \frac{1}{\pi}\int_{s_0}^{\infty} \frac{
\rho_{\mu}^{h}(s) }{s-p^2}ds,
\end{eqnarray}
where  $\rho_{\mu}^{h}$ is the spectral density of the higher
resonances and continuum. This spectral density can be approximated
by evoking the quark-–hadron duality assumption as:
\begin{eqnarray}\label{eq26}
\rho^h_{\mu}(s)&\simeq&\rho^{QCD}_{\mu}(s)\theta(s-s_0),
\end{eqnarray}
$\rho^{QCD}_{\mu}(s)$ is the perturbative  QCD spectral density
investigated from the theoretical side of the correlation function.
The threshold $s_0$ is chosen near the squared mass of the lowest
$D^0$ meson state.

Now, the QCD or the theoretical part of  the correlation function
should be calculated. The calculation of the  $\Pi_{\mu}$ in the
region of large space--like momentum  is based on the expansion of
the ${\cal T}$-product of the  interpolating and interaction
currents near the light-cone. After contracting $c$ and $\bar{c}$
quark fields, we get
\begin{eqnarray}\label{eq31}
\Pi_{\mu} &=& \int d^4x\, e^{iqx} \langle a_1^{-} (p', \varepsilon)
| \bar{d}(x)\, \gamma_\mu (1-\gamma_5) S^{c}(x,0) \gamma_5\,u(0) | 0
\rangle,
\end{eqnarray}
where  $S^{c}(x,0)$ is the full propagator of the $c$ quark in
presence of the background gluon field as
\begin{eqnarray}\label{eq32}
\,S^{c}(x)&=& \int \frac{d^4k}{(2\pi)^4} e^{-ikx} \frac{\not\!k +
m_c}{k^2-m_c^2}-g_s\int
\frac{d^4k}{(2\pi)^4}e^{-ikx}\int_0^1du\left[\frac{1}{2}\frac{k\!\!\!/+m_c}{(m_c^2-k^2)^2}G_{\mu\nu}(ux)\sigma^{\mu\nu}
\right.\nonumber\\
&+&\left.\frac{1}{m_c^2-k^2}ux_\mu G^{\mu\nu}(ux)\gamma_\nu\right].
\end{eqnarray}
The first term on the right-hand-side corresponds to the free quark
propagator, $G_{\mu\nu}$ is the gluon field strength tensor and
$g_s$ is the strong coupling constant. In the present work,
contributions with two gluons as well as four quark operators are
neglected because their contributions are small. For obtaining the
theoretical part  of the correlation function, the  Fierz
rearrangement is used. For this aim, the combination of
$\Gamma^{i}\Gamma_{i}$ is inserted before $u(0)$ in Eq.
(\ref{eq31}), where $\Gamma_{i}$ is the full set of the Dirac
matrices, $\Gamma_{i}= (I,~\gamma_5,~\gamma_\mu,~\gamma_{\mu}
\gamma_5,~\sigma_{\mu\nu})$. After rearrangement the quantum fields
and matrices appearing in the correlation function, in addition
considering all terms of the full propagator $S^{c}(x,0)$, it turns
into two parts including a matrix trace and a matrix element of
non--local operators between $a_1^{-}$ meson and vacuum state, i.e.,
$\langle a_1^{-}|\bar{d}(x) \,\Gamma_{i}\,u(0)|0\rangle$ and
$\langle a_1^{-}|\bar{d}(x) \,\Gamma_{i}\,
G_{\mu\nu}\,u(0)|0\rangle$. In the LCSR approach the non-zero matrix
elements, called the LCDAs, are defined in terms of twist functions.
For instance, two--particle DA $\langle a_1^{-}(p',\varepsilon) |
\bar{d}_{\alpha}(x)\, u_\delta (0) | 0 \rangle$ is presented as
\cite{Kwei}:
\begin{eqnarray}\label{eq34}
\langle a_1^{-}(p',\varepsilon) | \bar{d}_{\alpha}(x)\,u_\delta (0)
| 0 \rangle &=& -\frac{i}{4}  \int_0^1 du~ e^{i u  p'. x}\Bigg\{
f_{a_1^{-}} m_{a_1^{-}} \Bigg[ \not\! p'\gamma_5
\frac{\varepsilon^*. x}{p'.x} \Phi_\parallel(u) +\Bigg( \not\!
\varepsilon^* -\not\! p'
\frac{\varepsilon^*. x}{p'.x}\Bigg)\gamma_5 g_\perp^{(a)}(u) \nonumber\\
&-& \not\! x\gamma_5 \frac{\varepsilon^*. x}{2(p'.x)^2} m_{a_1}^2
\phi_{b}(u) + \epsilon_{\mu\nu\rho\sigma} \varepsilon^{*\nu}
p'^{\rho} x^\sigma \gamma^\mu
\frac{g_\perp^{(v)}(u)}{4}\Bigg] \nonumber\\
&+& \,f^{\perp}_{A} \Bigg[ \frac{1}{2}( \not\! p'\not\!\epsilon^*-
\not\!\epsilon^* \not\! p' ) \gamma_5\,\Phi_\perp(u) - \frac{1}{2}(
\not\! p'\not\! x- \not\! x \not\! p' ) \gamma_5 \frac{\epsilon^*.
x}{(p'.x)^2} m_{a_1}^2 \bar
h_\parallel^{(t)} (u) \nonumber\\
&+& i \Big(\epsilon^*. x\Big) m_{a_1}^2 \gamma_5
\frac{h^{(p)}_\parallel (u)}{2} \Bigg]\Bigg\}_{\delta\alpha},
\end{eqnarray}
where $\Phi_\parallel$, $\Phi_\perp$ are twist-2, $g_\perp^{(a)}$,
$g_\perp^{(v)}$, $h_\parallel^{(t)}$ and $h_\parallel^{(p)}$ are
twist-3  functions. For $x^{2}\neq 0$, we have
\begin{eqnarray*}\label{eq35}
\phi_{b} (u) &=&\Phi_\parallel -2 g_\perp^{(a)}(u),\nonumber\\
\bar h_\parallel^{(t)} &=& h_\parallel^{(t)}- \frac{1}{2}
\Phi_\perp(u).
\end{eqnarray*}
We should keep out all terms of the two--parton LCDA in Eq.
(\ref{eq34}) in our calculations, since the mass of the axial vector
meson $a_1^{-}$ is comparable to the charm quark mass. The explicit
expressions for the relevant two-- and three--parton LCDAs and
definitions for the above  mentioned twist functions are collected
in Appendix \ref{app:fun-def}.

Using the LCDAs and after some straightforward calculations, the
correlation function in theoretical side appears as an integral
expression that made up of the twist functions and Lorentz
structures.

To equate the coefficients of the corresponding Lorentz structures
from both phenomenological and theoretical sides of the correlation
function and apply Borel transform with respect to the variable
$p^2$ as
\begin{eqnarray}\label{eq316}
B_{p^2}(M^2)\frac{1}{\left(
p^{2}-m_{D^0}^{2}\right)^{n}}&=&\frac{(-1)^{n}}{\Gamma(n)}\frac{e^{-\frac{m_{D^0}^{2}}{M^{2}}}}{(M^{2})^{n}},
\end{eqnarray}
in order to suppress the higher states and continuum contributions, one can obtain the transition form factors of the $D^0 \to a_1^{-}
\ell^+ \nu$ decay in the frame work of the LCSR. For instance, the
form factor $A(q^{2})$ is calculated as
\begin{eqnarray}\label{eq317}
A(q^{2})&=&-
\frac{m_{c}(m_{D^0}-m_{a_1^{-}})\,f_{a_1^{-}}}{m_{D^0}^2\,f_{D^0}}\,e^{m_{D^0}^2/M^2}\int_{u_0}^{1}du\,
e^{s(u)}\Bigg[ \frac{9
f_{a_1^{-}}^{\perp}}{f_{a_1^{-}}}\,\frac{\Phi_\perp (u)}{u}
+\frac{32 f_{a_1^{-}}^{\perp}}{f_{a_1^{-}}}\frac{
m_{a_1^{-}}^{2}}{M^2}~\frac{\bar{h}{_\parallel^{(t)(ii)}(u)}}{u}+\frac{m_{c}m_{a_1^{-}}}{2\,M^{2}}
\frac{{g_\perp^{(v)}(u)}}{u^2} \Bigg],\nonumber\\
\end{eqnarray}
where $u_{0}$ is the function of $s_0$, the continuum
threshold of $D^0$ meson, as
\begin{eqnarray}\label{eq317a}
u_{0}(s_0) =\frac{1}{2m^2_{a_1^{-}}}
\left[\sqrt{(s_0-m_{a_1^{-}}^2-q^2)^2 +4 m_{a_1^{-}}^2 (m_c^2-q^2)}
-(s_0-m_{a_1^{-}}^2-q^2)\right].
\end{eqnarray}
The explicit expressions for the other form factors are presented in
Appendix \ref{app:form factors}.

Following the previous steps in this section, phrases similar to Eq.
(\ref{eq317}) and Appendix \ref{app:form factors} can be obtained
for the transition form factors of $D^{0} \to b^{-}_{1} \ell^+ \nu$,
$D^{+} \to a^{0}_{1} (b^{0}_{1}) \ell^+ \nu$, $D_{s}^{+} \to
K^{0}_{1} \ell^+ \nu$ as well as $D^{+}\to K^{0}_{1} \ell^+ \nu$
decays via the LCSR approach.

\section{Numerical analysis}\label{se.nu}

We present our numerical analysis for the form factors and branching
ratio values of the semileptonic $D_{(s)} \to A \ell^+ \nu$, where
$A=a_{1}, b_{1}, K_{1}(1270,1400)$, and the nonleptonic $D \to K_1
(1270,1400) \pi$ decays in two subsections.  First, the transition
form factors and branching ratio values of the semileptonic $D^{+}
\to a^{0}_{1} (b^{0}_{1}, K^{0}_{1}) \ell^+ \nu$, $D^{0} \to
a^{-}_{1} (b^{-}_{1}) \ell^+ \nu$, and $D^{+}_{s}\to K^{0}_{1}
\ell^+ \nu$ decays are analyzed. In the next subsection, using these
form factors, the branching ratio values are calculated for the
nonleptonic $D^0 \to K_1^{-}(1270) \pi^+$, $D^0 \to K_1^{-}(1400)
\pi^+$, $D^+ \to K_1^0 (1270) \pi^+$ and $D^+ \to K_1^0 (1400)
\pi^+$ decays via the factorization method. For a better analysis, a
comparison is made between our results and predictions of the other
methods and the experimental values.

In this work, masses  are taken in $\mbox{GeV}$ as $m_c=1.28\pm
0.03$, $m_{D}=1.86  $ and $m_{D_s}= 1.96 $ \cite{pdg}. We use the
results of the QCD sum rules for decay constants of $D$ and $D_s$
mesons, rather than the actual value of them, as $f_{D}=210 \pm
{12}~\mbox{MeV}$ and $f_{D_s}= 246\pm 8~\mbox{MeV}$ \cite{Mutuk}; in
this way radiative correction will be canceled. Masses and decay
constant values for the axial vector mesons are collected in Table
\ref{Ta1}.  We can take $f_{A}=f^\perp_{A}$ at energy scale
$\mu=1\,\rm{GeV}$ \cite{Kwei}. All of the decay constant values for
the axial vector mesons in Table \ref{Ta1}, and also masses for two
$K_{1A}$ and $K_{1B}$ states are estimated from the LCSR
\cite{Kwei}.
\begin{table}[th]
\caption{Masses and decay constants for axial vector mesons and two
states $K_{1A}$ and $K_{1B}$ \cite{pdg,Kwei}.} \label{Ta1}
\begin{ruledtabular}
\begin{tabular}{ccccc}
Mass &$m_{a_1}$&$m_{b_1}$&$m_{K_{1A}}$&$m_{K_{1B}}$\\
\hline
Value\,(GeV)&$1.23\pm0.40$&$1.23\pm0.32$&$1.31\pm0.06$&$1.34\pm0.08$\\
\hline\hline
Decay Constant&$f_{a_1}$&$f_{b_1}$&$f_{K_{1A}}$&$f_{K_{1B}}$\\
\hline
Value\,(MeV)&$238\pm10$&$180\pm8$&$250\pm13$&$190\pm10$\\
\end{tabular}
\end{ruledtabular}
\end{table}

It should be noted that the decay constants of $K_1(1270)$ and
$K_1(1400)$ mesons are written in terms of $f_{K_{1A}}$ and
$f_{K_{1B}}$ as \cite{Kwei}:
\begin{eqnarray}\label{eq301}
f_{K_1(1270)} &=&\sin {\theta_K}\, \frac{ m_{K_{1A}}}{
m_{K_1(1270)}} f_{K_{1A}}+ \cos {\theta_K}\,\frac{ m_{K_{1B}}}{
m_{K_1(1270)}}
a_0^{\parallel,K_{1B}} f_{K_{1B}}, \nonumber\\
f_{K_1(1400)} &=&\cos {\theta_K}\, \frac{ m_{K_{1A}}}{
m_{K_1(1400)}} f_{K_{1A}}- \sin {\theta_K}\,\frac{ m_{K_{1B}}}{
m_{K_1(1400)}} a_0^{\parallel,K_{1B}} f_{K_{1B}},
\end{eqnarray}
where $a_0^{\parallel,K_{1B}}$ is G-parity invariant Gegenbauer
moment for  $K_{1B}$ state.

\subsection{Analysis of semileptonic decays }

From the formulas presented in Eq. (\ref{eq317}) and Appendix
\ref{app:form factors} for the form factors of the semileptonic $D^0
\to a_1^{-} \ell^+ \nu$ decays, it is easily known that they contain
two free parameters $M^{2}$ and $s_{0}$, which are the Borel
mass--square and the continuum threshold of $D^0$ meson,
respectively. $M^2$ and $s_{0}$ are not physical quantities;
therefore,  the form factors as physical quantities should be
independent of them. The Borel parameter must be large enough to
suppress  the contribution of higher states. On the other hand, it
should be small enough  to show the effect of twist functions.

\begin{figure}
\includegraphics[width=6cm,height=5.75cm]{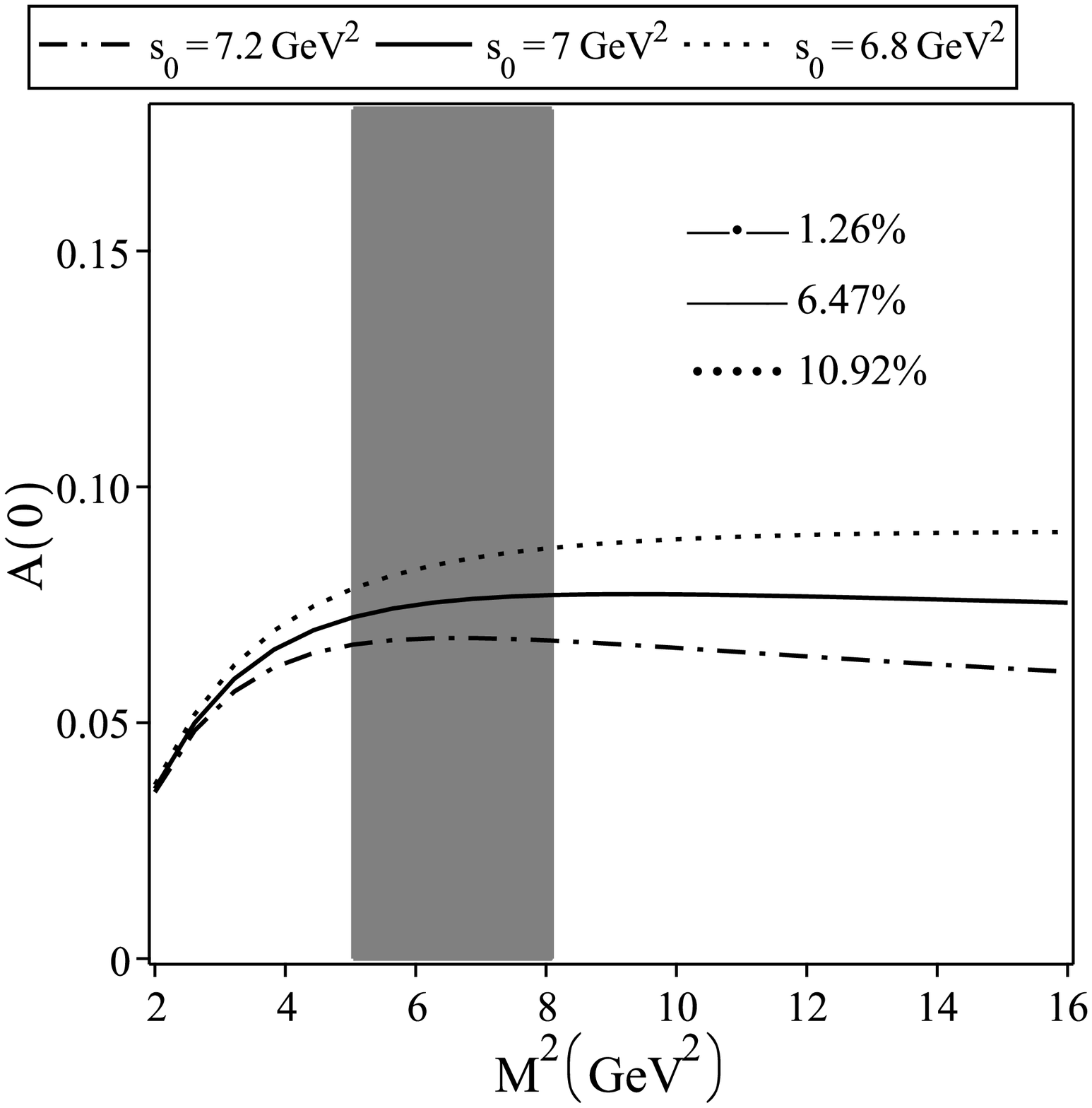}
\includegraphics[width=6cm,height=5.75cm]{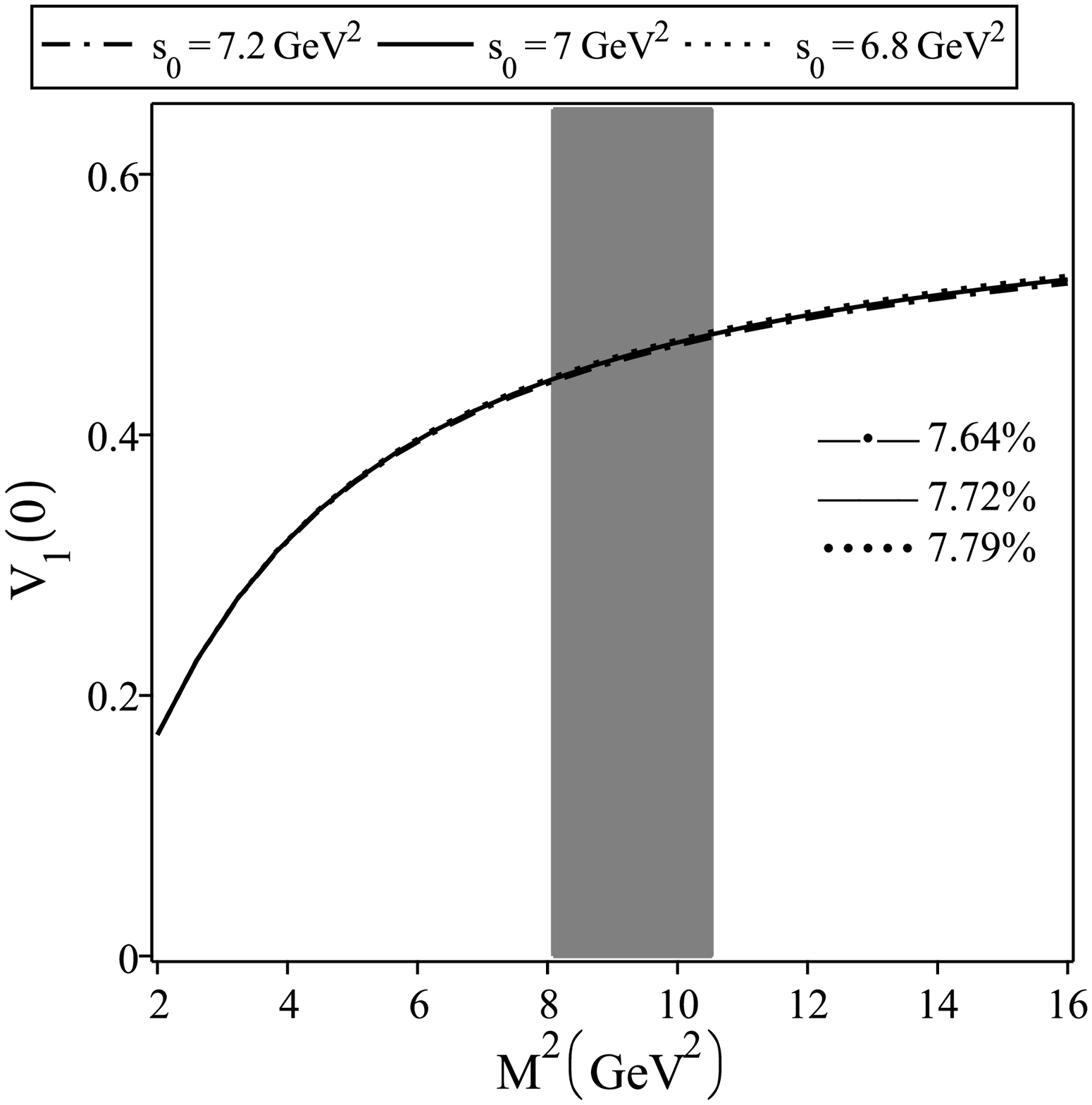}
\includegraphics[width=6cm,height=5.75cm]{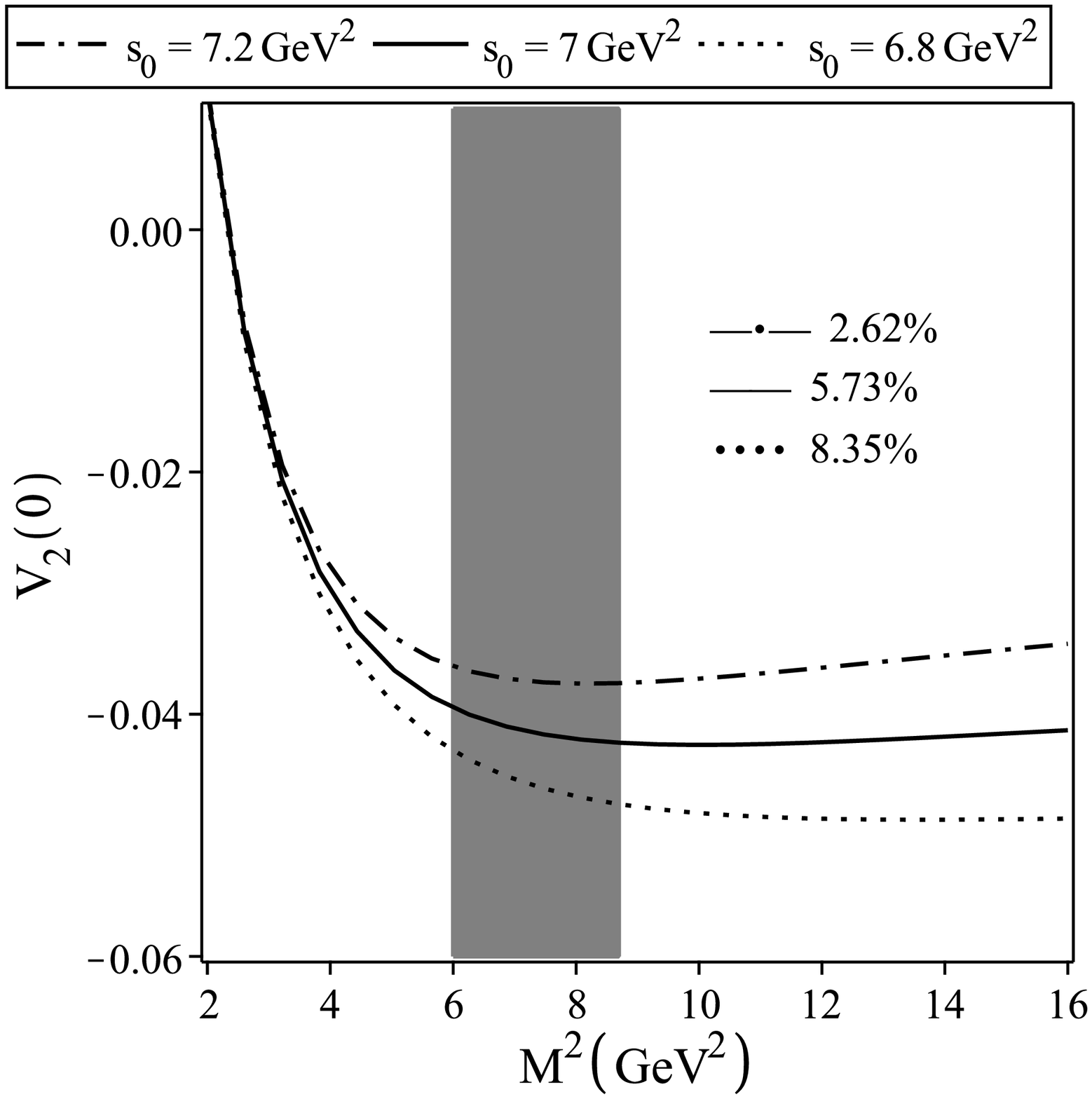}
\includegraphics[width=6cm,height=5.75cm]{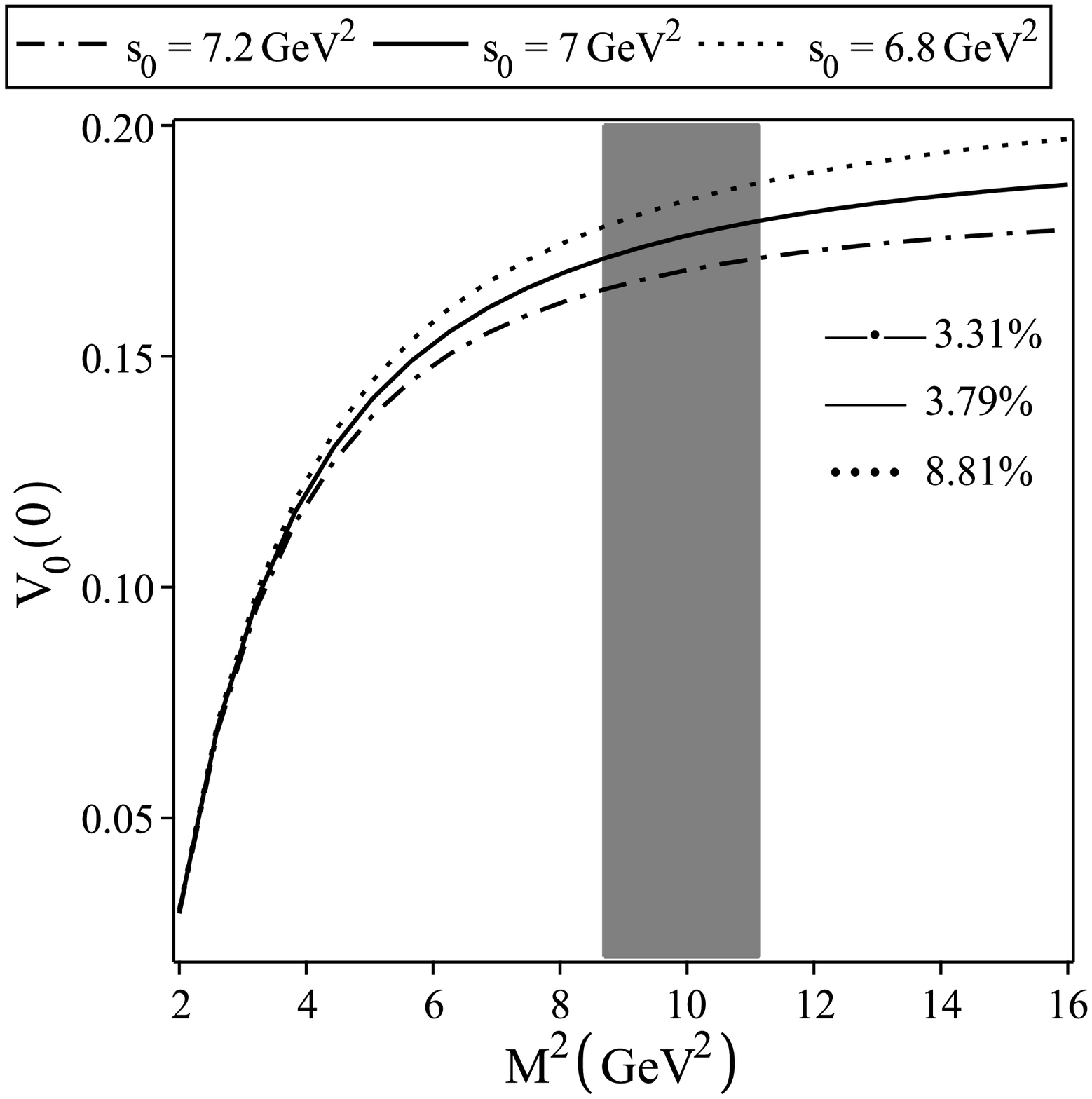}
\caption{$D^0 \to a_1^{-}$ transition form factors as functions of
$M^{2}$ at $q^2=0$. The dot, solid and dot-dashed lines correspond
to $s_{0}=6.8,~ 7$ and $7.2~\mbox{GeV}^2$, respectively}\label{Fb}
\end{figure}

For instance in Fig. \ref{Fb}, the dependence of the $D^0 \to
a_1^{-}$ form factors is displaced with respect to $M^{2}$, at
$q^2=0$, for three values of $s_{0}=6.8,~7$ and $7.2~ \mbox{GeV}^2$
with dot, solid and dash-dot lines,  respectively. In this figure,
shaded interval shows the proper region of the Borel parameter for
each transition form factor of the semileptonic $D^0 \to a_1^{-}$
decay.
As can be seen in Fig. \ref{Fb}, the form factors $A$, $V_{1}$,
$V_{2}$ and $V_{0}$ of the $D^0 \to a_1^{-}$ transition, obtained
from the sum rules in $s_{0}=7~\mbox{GeV}^2$, can be stable within
the Borel mass intervals  $5~ \mbox{GeV}^2 <M^{2}<8~ \mbox{GeV}^2$,
$ 8~ \mbox{GeV}^2 <M^{2}< 10~ \mbox{GeV}^2$, and $6 ~ \mbox{GeV}^2
<M^{2}< 9~\mbox{GeV}^2$ and $9~\mbox{GeV}^2 <M^{2}<
11~\mbox{GeV}^2$, respectively. From now on, we get the continuum
threshold of $D_{(s)}$ meson for all decays equal to
$s_{0}=7~\mbox{GeV}^2$ in our calculations.

Having all the required parameters, we can estimate the form factors
for each aforementioned semileptonic decay. The LCSR predictions for
the form factors are valid in half of the physical region $ m_\ell^2
\le q^2 \le (m_{D_{(s)}}-m_{A})^2$, nearly, and then these
quantities are truncated at some points. In order to extend our
results to the full physical region, we look for parametrization of
the form factors in such a way that in the validity region of the
LCSR, this parametrization coincides with the sum rules predictions.
We use the following fit functions of the form factors with respect
to $q^2$ as:
\begin{eqnarray}\label{eq411}
F^{(1)}(q^2)&=&\frac{1}{1- {q^2}/{m_{D_{(s)}}^2}}\sum_{l=0}^{2} b_l
\left[z^l + (-1)^{l}\, \frac{l}{3}\, z^4 \right]\,,
\\
F^{(2)}(q^{2})&=&\frac{F(0)}{1-\alpha\,({q^2}/{m_{D_{(s)}}^2})+\beta\,{({q^2}/{m_{D_{(s)}}^2})}^2}\,,
\\
F^{(3)}(q^{2})&=&\frac{c_1}{1-q^2/m_{D^{*}}^{2}}+\frac{c_2}{(1-q^2/m_{D^{*}}^{2})^{2}}\,,
\\
F^{(4)}(q^{2})&=&\frac{r_{1}}{1-q^2/m_{D^{*}}^{2}}+\frac{r_{2}}{1-q^2/m_{\rm{fit}}^{2}}\,,
\end{eqnarray}
where
$z=\frac{\sqrt{t_{+}-q^2}-\sqrt{t_{+}-t_{0}}}{\sqrt{t_{+}-q^2}+\sqrt{t_{+}-t_{0}}}$,
$t_{+}=(m_{D_{(s)}}+m_{A})^2$ and
$t_{0}=(m_{D_{(s)}}+m_{A})(\sqrt{m_{D_{(s)}}}-\sqrt{m_{A}})^2$
\cite{Bourrely}.

Tables \ref{T331}, \ref{T33}, \ref{T333} and \ref{T3333} show the
values of [$b_{0},~ b_{1},~ b_{2}$],  $[F(0),~ \alpha,~  \beta]$,
$[c_1,~ c_2]$, and also $[r_{1},~ r_{2},~ m_{\rm{fit}}]$
respectively, for the form factors of the semileptonic decays.
\begin{table}[th]
\caption{Values of $b_{0}$, $b_{1}$ and $b_{2}$ related to
$F^{(1)}(q^2)$ for the fitted form factors of $D_{(s)}\to a_{1},
b_{1}, K_{1A}$ and $K_{1B}$ transitions.} \label{T331}
\begin{ruledtabular}
\begin{tabular}{cccc|cccc}
$\mbox{Form factor}$& ${b_{0}}$&$b_{1}$&$b_{2}$ &$\mbox{Form factor}$& ${b_{0}}$&$b_{1}$&$b_{2}$\\
\hline
$A^{D^0\to a^{-}_{1}}$& ${0.08}$&$0.45$&$19.42$ &$A^{D^0\to b^{-}_{1}}$& ${-0.39}$&$-2.75$&$25.99$\\
$V_{1}^{D^0\to a^{-}_{1}}$& ${0.40}$&$3.49$&$6.64$ &$V_{1}^{D^0\to b^{-}_{1}}$& ${-0.22}$&$0.07$&$36.37$\\
$V_{2}^{D^0\to a^{-}_{1}}$& ${-0.03}$&$-0.74$&$-1.21$ &$V_{2}^{D^0\to b^{-}_{1}}$& ${0.19}$&$2.72$&$23.43$\\
$V_{0}^{D^0\to a^{-}_{1}}$& ${0.14}$&$0.99$&$-9.16$ &$V_{0}^{D^0\to b^{-}_{1}}$& ${-0.25}$&$-10.98$&$-64.47$\\
$A^{D^+\to a^{0}_{1}}$& ${0.05}$&$0.27$&$12.35$ &$A^{D^+\to b^{0}_{1}}$& ${-0.26}$&$-1.94$&$20.47$\\
$V_{1}^{D^+\to a^{0}_{1}}$& ${0.27}$&$2.38$&$4.01$ &$V_{1}^{D^+\to b^{0}_{1}}$& ${-0.16}$&$0.09$&$25.63$\\
$V_{2}^{D^+\to a^{0}_{1}}$& ${-0.03}$&$-0.56$&$-0.71$ &$V_{2}^{D^+\to b^{0}_{1}}$& ${0.13}$&$1.96$&$16.72$\\
$V_{0}^{D^+\to a^{0}_{1}}$& ${0.69}$&$0.71$&$-5.63$ &$V_{0}^{D^+\to b^{0}_{1}}$& ${-0.23}$&$0.49$&$217.83$\\
$A^{D\to K_{1A}}$& ${0.06}$&$0.49$&$18.92$ &$A^{D\to K_{1B}}$& ${-0.51}$&$-3.65$&$39.48$\\
$V_{1}^{D\to K_{1A}}$& ${0.35}$&$3.40$&$5.02$ &$V_{1}^{D\to K_{1B}}$& ${-0.29}$&$0.14$&$66.82$\\
$V_{2}^{D\to K_{1A}}$& ${-0.02}$&$-0.59$&$0.38$ &$V_{2}^{D\to K_{1B}}$& ${0.29}$&$5.31$&$49.19$\\
$V_{0}^{D\to K_{1A}}$& ${0.16}$&$2.82$&$-116.30$ &$V_{0}^{D\to K_{1B}}$& ${0.36}$&$-13.69$&$-119.16$\\
$A^{D_s\to K_{1A}}$& ${0.06}$&$0.39$&$18.73$ &$A^{D_s\to K_{1B}}$& ${-0.47}$&$-3.11$&$37.49$\\
$V_{1}^{D_s\to K_{1A}}$& ${0.32}$&$2.83$&$3.96$ &$V_{1}^{D_s\to K_{1B}}$& ${-0.27}$&$0.35$&$56.71$\\
$V_{2}^{D_s\to K_{1A}}$& ${-0.01}$&$-0.40$&$-0.64$ &$V_{2}^{D_s\to K_{1B}}$& ${0.26}$&$4.61$&$45.89$\\
$V_{0}^{D_s\to K_{1A}}$& ${0.09}$&$0.54$&$-10.98$ &$V_{0}^{D_s\to K_{1B}}$& ${-0.33}$&$-16.04$&$-166.56$
\end{tabular}
\end{ruledtabular}
\end{table}
\begin{table}[th]
\caption{Values of $F(0)$, $\alpha$ and $\beta$  connected to
$F^{(2)}(q^2)$ for the fitted form factors of $D_{(s)}\to a_{1},
b_{1}, K_{1A}$ and $K_{1B}$ transitions.} \label{T33}
\begin{ruledtabular}
\begin{tabular}{cccc|cccc}
$\mbox{Form factor}$& ${F(0)}$&$\alpha$&$\beta$ &$\mbox{Form factor}$& ${F(0)}$&$\alpha$&$\beta$\\
\hline
$A^{D^0\to a^{-}_{1}}$& ${0.08}$&$0.25$&$-2.17$ &$A^{D^0\to b^{-}_{1}}$& ${-0.41}$&$0.45$&$0.40$\\
$V_{1}^{D^0\to a^{-}_{1}}$& ${0.42}$&$0.23$&$-0.18$ &$V_{1}^{D^0\to b^{-}_{1}}$& ${-0.22}$&$1.20$&$1.47$\\
$V_{2}^{D^0\to a^{-}_{1}}$& ${-0.04}$&$-0.72$&$1.54$ &$V_{2}^{D^0\to b^{-}_{1}}$& ${0.21}$&$-0.31$&$-0.38$\\
$V_{0}^{D^0\to a^{-}_{1}}$& ${0.15}$&$0.46$&$0.36$ &$V_{0}^{D^0\to b^{-}_{1}}$& ${-0.32}$&$-2.30$&$9.12$\\
$A^{D^+\to a^{0}_{1}}$& ${0.05}$&$0.26$&$-2.21$ &$A^{D^+\to b^{0}_{1}}$& ${-0.28}$&$0.44$&$0.49$\\
$V_{1}^{D^+\to a^{0}_{1}}$& ${0.29}$&$0.24$&$-0.17$ &$V_{1}^{D^+\to b^{0}_{1}}$& ${-0.16}$&$1.22$&$1.45$\\
$V_{2}^{D^+\to a^{0}_{1}}$& ${-0.03}$&$-0.72$&$1.61$ &$V_{2}^{D^+\to b^{0}_{1}}$& ${0.15}$&$-0.32$&$-0.36$\\
$V_{0}^{D^+\to a^{0}_{1}}$& ${0.10}$&$0.41$&$0.34$ &$V_{0}^{D^+\to b^{0}_{1}}$& ${-0.23}$&$-2.15$&$8.47$\\
$A^{D\to K_{1A}}$& ${0.07}$&$0.21$&$-2.14$ &$A^{D\to K_{1B}}$& ${-0.53}$&$0.46$&$0.38$\\
$V_{1}^{D\to K_{1A}}$& ${0.37}$&$0.20$&$-0.13$ &$V_{1}^{D\to K_{1B}}$& ${-0.29}$&$1.17$&$1.72$\\
$V_{2}^{D\to K_{1A}}$& ${-0.03}$&$-0.70$&$1.81$ &$V_{2}^{D\to K_{1B}}$& ${0.31}$&$-0.49$&$-0.21$\\
$V_{0}^{D\to K_{1A}}$& ${0.11}$&$0.44$&$0.61$ &$V_{0}^{D\to K_{1B}}$& ${-0.42}$&$-2.33$&$9.24$\\
$A^{D_s\to K_{1A}}$& ${0.07}$&$0.25$&$-2.34$ &$A^{D_s\to K_{1B}}$& ${-0.49}$&$0.50$&$0.44$\\
$V_{1}^{D_s\to K_{1A}}$& ${0.34}$&$0.24$&$-0.15$ &$V_{1}^{D_s\to K_{1B}}$& ${-0.27}$&$1.28$&$1.80$\\
$V_{2}^{D_s\to K_{1A}}$& ${-0.02}$&$-0.84$&$1.92$ &$V_{2}^{D_s\to K_{1B}}$& ${0.29}$&$-0.53$&$-0.24$\\
$V_{0}^{D_s\to K_{1A}}$& ${0.10}$&$0.61$&$0.72$ &$V_{0}^{D_s\to K_{1B}}$& ${-0.41}$&$-2.37$&$9.92$
\end{tabular}
\end{ruledtabular}
\end{table}
\begin{table}[th]
\caption{Values of $c_{1}$ and $c_{2}$ connected to $F^{(3)}(q^2)$
for the fitted form factors of $D_{(s)}\to a_{1}, b_{1}, K_{1A}$ and
$K_{1B}$ transitions.} \label{T333}
\begin{ruledtabular}
\begin{tabular}{ccc|ccc}
$\mbox{Form factor}$& ${c_{1}}$&$c_{2}$ &$\mbox{Form factor}$& ${c_{1}}$&$c_{2}$\\
\hline
$A^{D^0\to a^{-}_{1}}$& ${0.12}$&$-0.04$ &$A^{D^0\to b^{-}_{1}}$& ${-0.63}$&$0.22$\\
$V_{1}^{D^0\to a^{-}_{1}}$& ${0.72}$&$-0.30$ &$V_{1}^{D^0\to b^{-}_{1}}$& ${-0.19}$&$-0.02$\\
$V_{2}^{D^0\to a^{-}_{1}}$& ${-0.10}$&$0.06$ &$V_{2}^{D^0\to b^{-}_{1}}$& ${0.45}$&$-0.24$\\
$V_{0}^{D^0\to a^{-}_{1}}$& ${0.23}$&$-0.08$ &$V_{0}^{D^0\to b^{-}_{1}}$& ${-1.30}$&$0.98$\\
$A^{D^+\to a^{0}_{1}}$& ${0.07}$&$-0.02$ &$A^{D^+\to b^{0}_{1}}$& ${-0.43}$&$0.15$\\
$V_{1}^{D^+\to a^{0}_{1}}$& ${0.49}$&$-0.20$ &$V_{1}^{D^+\to b^{0}_{1}}$& ${-0.13}$&$-0.02$\\
$V_{2}^{D^+\to a^{0}_{1}}$& ${-0.07}$&$0.04$ &$V_{2}^{D^+\to b^{0}_{1}}$& ${0.32}$&$-0.18$\\
$V_{0}^{D^+\to a^{0}_{1}}$& ${0.15}$&$-0.05$ &$V_{0}^{D^+\to b^{0}_{1}}$& ${-0.07}$&$-0.15$\\
$A^{D\to K_{1A}}$& ${0.11}$&$-0.04$ &$A^{D\to K_{1B}}$& ${-0.81}$&$0.28$\\
$V_{1}^{D\to K_{1A}}$& ${0.65}$&$-0.28$ &$V_{1}^{D\to K_{1B}}$& ${-0.26}$&$-0.02$\\
$V_{2}^{D\to K_{1A}}$& ${-0.07}$&$0.04$ &$V_{2}^{D\to K_{1B}}$& ${0.74}$&$-0.43$\\
$V_{0}^{D\to K_{1A}}$& ${0.17}$&$-0.06$ &$V_{0}^{D\to K_{1B}}$& ${0.01}$&$-0.43$\\
$A^{D_s\to K_{1A}}$& ${0.11}$&$-0.04$ &$A^{D_s\to K_{1B}}$& ${-0.73}$&$0.24$\\
$V_{1}^{D_s\to K_{1A}}$& ${0.58}$&$-0.24$ &$V_{1}^{D_s\to K_{1B}}$& ${-0.21}$&$-0.05$\\
$V_{2}^{D_s\to K_{1A}}$& ${-0.05}$&$0.03$ &$V_{2}^{D_s\to K_{1B}}$& ${0.69}$&$-0.40$\\
$V_{0}^{D_s\to K_{1A}}$& ${0.14}$&$-0.04$ &$V_{0}^{D_s\to K_{1B}}$&
${-1.82}$&$1.41$
\end{tabular}
\end{ruledtabular}
\end{table}
\begin{table}[th]
\caption{Values of $r_{1}$, $r_{2}$ and $m_{\rm{fit}}$  connected to
$F^{(4)}(q^2)$ for the fitted form factors of $D_{(s)}\to a_{1},
b_{1}, K_{1A}$ and $K_{1B}$ transitions.} \label{T3333}
\begin{ruledtabular}
\begin{tabular}{cccc|cccc}
$\mbox{Form factor}$& ${r_{1}}$&$r_{2}$&$m_{\rm{fit}}$ & $\mbox{Form factor}$& ${r_{1}}$&$r_{2}$&$m_{\rm{fit}}$\\
\hline
$A^{D^0\to a^{-}_{1}}$& ${1.34}$&$-1.26$&$1.82$ &$A^{D^0\to b^{-}_{1}}$& ${-1.86}$&$1.45$&$1.73$\\
$V_{1}^{D^0\to a^{-}_{1}}$& ${3.82}$&$-3.40$&$1.78$ &$V_{1}^{D^0\to b^{-}_{1}}$& ${1.49}$&$-1.71$&$1.84$\\
$V_{2}^{D^0\to a^{-}_{1}}$& ${-1.07}$&$1.03$&$1.80$ &$V_{2}^{D^0\to b^{-}_{1}}$& ${3.86}$&$-3.65$&$1.79$\\
$V_{0}^{D^0\to a^{-}_{1}}$& ${0.77}$&$-0.62$&$1.75$ &$V_{0}^{D^0\to b^{-}_{1}}$& ${-17.34}$&$17.02$&$1.80$\\
$A^{D^+\to a^{0}_{1}}$& ${0.91}$&$-0.86$&$1.82$ &$A^{D^+\to b^{0}_{1}}$& ${-1.71}$&$0.89$&$1.71$\\
$V_{1}^{D^+\to a^{0}_{1}}$& ${2.52}$&$-2.30$&$1.78$ &$V_{1}^{D^+\to b^{0}_{1}}$& ${1.13}$&$-1.29$&$1.84$\\
$V_{2}^{D^+\to a^{0}_{1}}$& ${-0.71}$&$0.68$&$1.79$ &$V_{2}^{D^+\to b^{0}_{1}}$& ${2.80}$&$-2.65$&$1.80$\\
$V_{0}^{D^+\to a^{0}_{1}}$& ${0.62}$&$-0.52$&$1.76$ &$V_{0}^{D^+\to b^{0}_{1}}$& ${9.37}$&$-9.60$&$1.84$\\
$A^{D\to K_{1A}}$& ${0.93}$&$-0.86$&$1.81$ &$A^{D\to K_{1B}}$& ${-2.04}$&$1.51$&$1.70$\\
$V_{1}^{D\to K_{1A}}$& ${2.91}$&$-2.54$&$1.76$ &$V_{1}^{D\to K_{1B}}$& ${1.15}$&$-1.44$&$1.84$\\
$V_{2}^{D\to K_{1A}}$& ${-0.63}$&$0.60$&$1.78$ &$V_{2}^{D\to K_{1B}}$& ${4.39}$&$-4.08$&$1.76$\\
$V_{0}^{D\to K_{1A}}$& ${0.26}$&$-0.15$&$1.57$ &$V_{0}^{D\to K_{1B}}$& ${9.35}$&$-9.77$&$1.82$\\
$A^{D_s\to K_{1A}}$& ${1.18}$&$-1.11$&$1.92$ &$A^{D_s\to K_{1B}}$& ${-1.42}$&$0.93$&$1.74$\\
$V_{1}^{D_s\to K_{1A}}$& ${2.85}$&$-2.48$&$1.87$ &$V_{1}^{D_s\to K_{1B}}$& ${2.12}$&$-2.39$&$1.93$\\
$V_{2}^{D_s\to K_{1A}}$& ${-0.47}$&$0.45$&$1.88$ &$V_{2}^{D_s\to K_{1B}}$& ${5.05}$&$-4.76$&$1.88$\\
$V_{0}^{D_s\to K_{1A}}$& ${0.13}$&$-0.03$&$1.36$ &$V_{0}^{D_s\to
K_{1B}}$& ${-20.99}$&$20.58$&$1.89$
\end{tabular}
\end{ruledtabular}
\end{table}

\begin{figure}
\includegraphics[width=6cm,height=6cm]{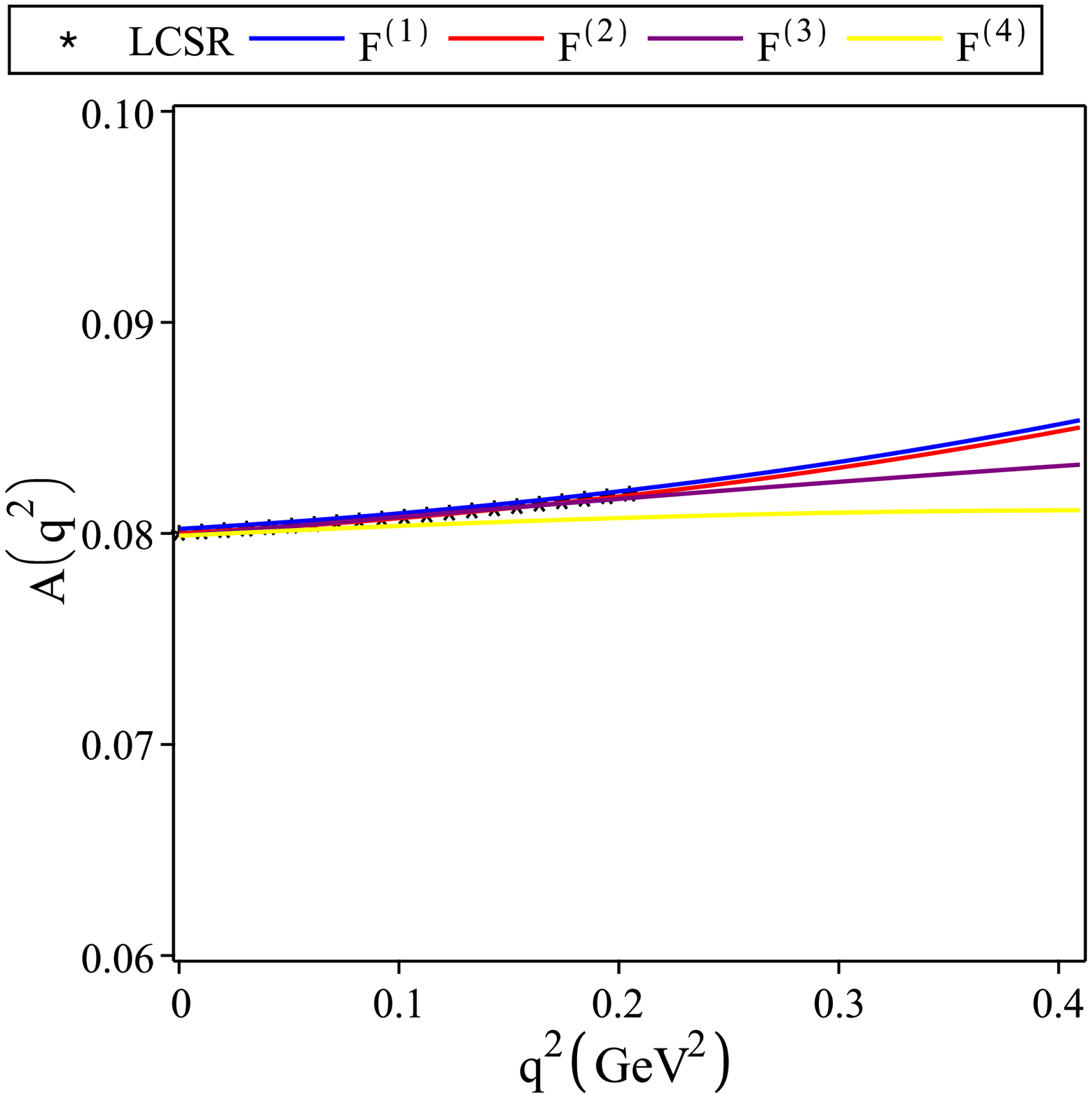}
\includegraphics[width=6cm,height=6cm]{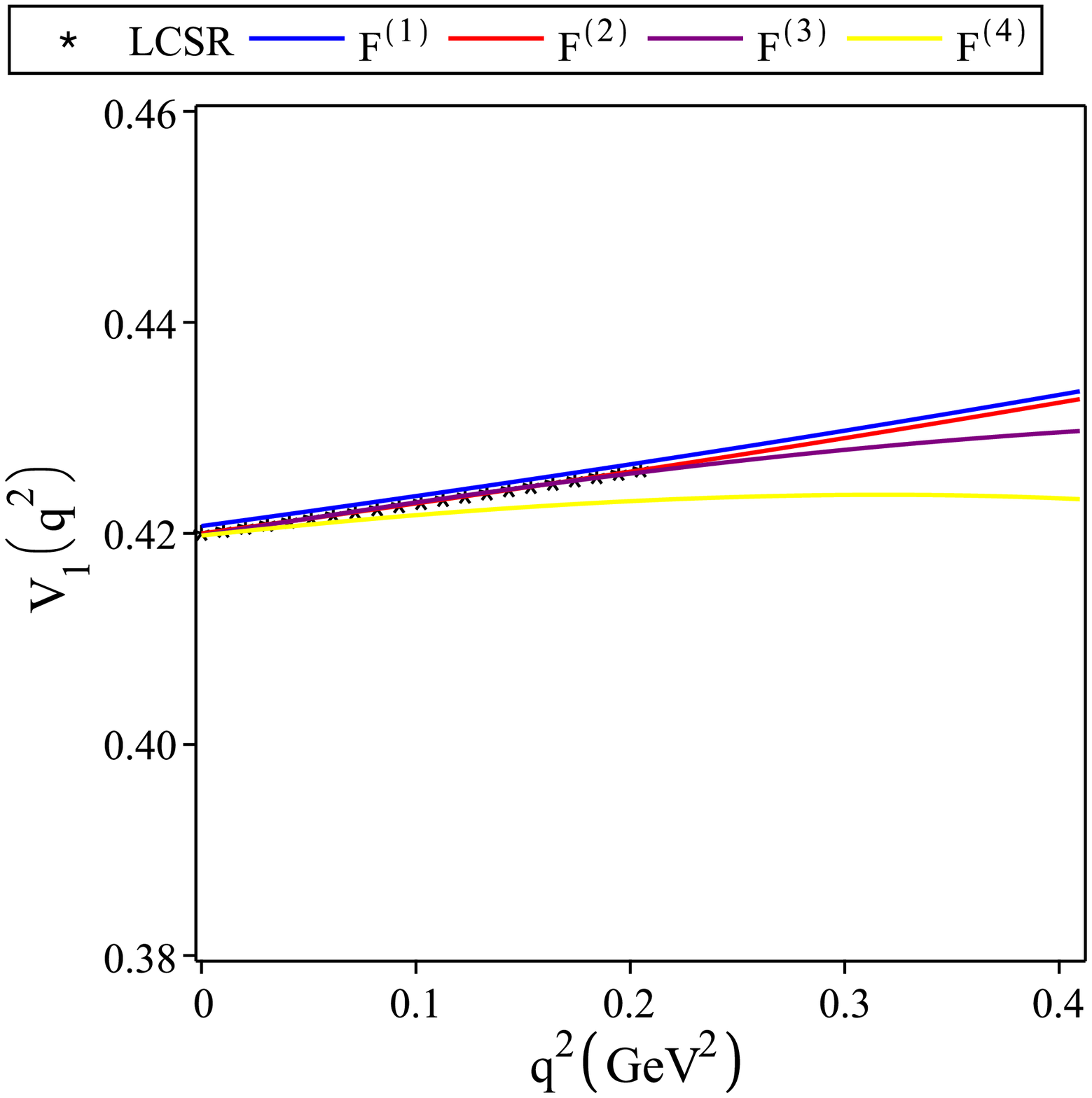}
\includegraphics[width=6cm,height=6cm]{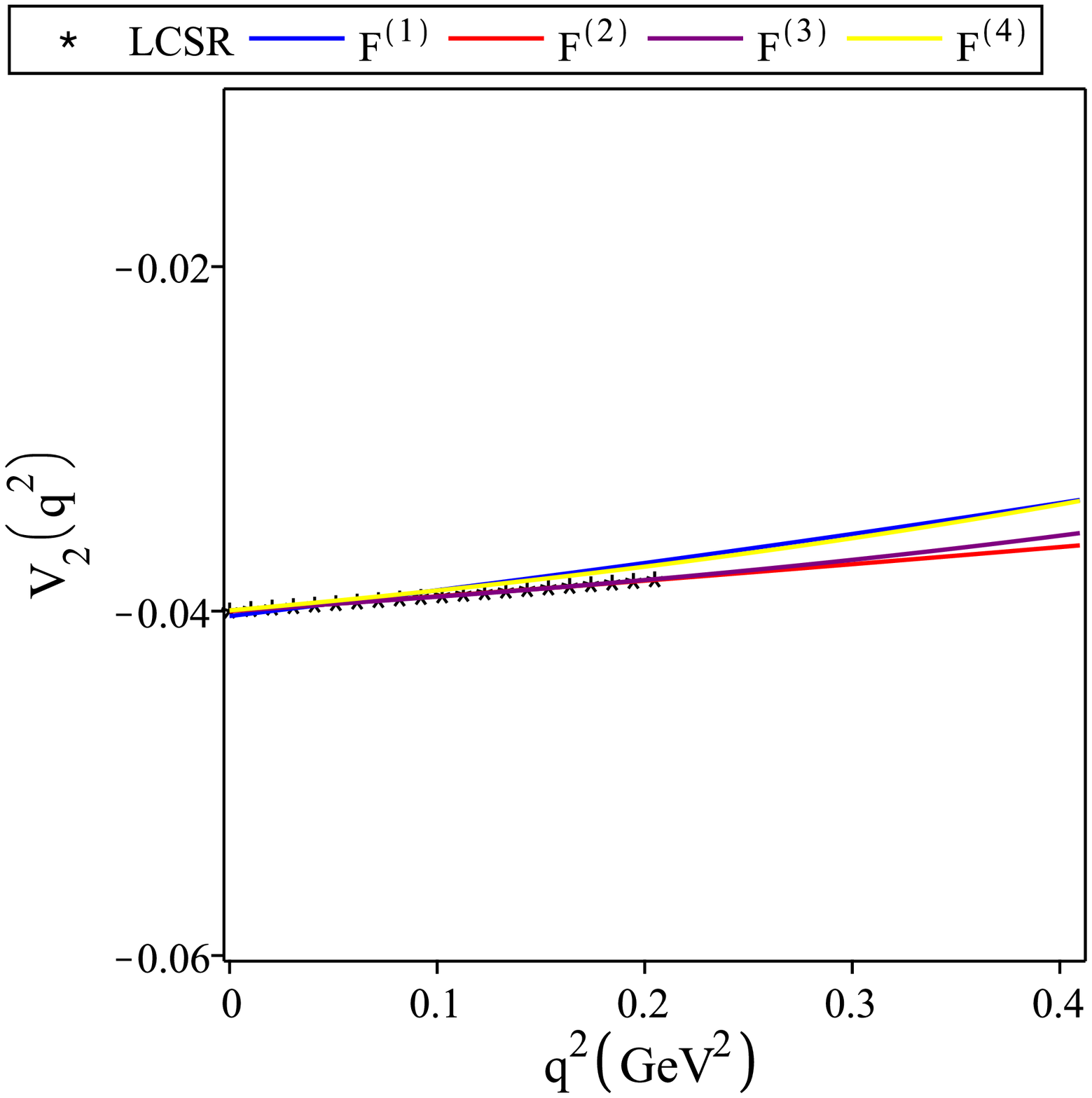}
\includegraphics[width=6cm,height=6cm]{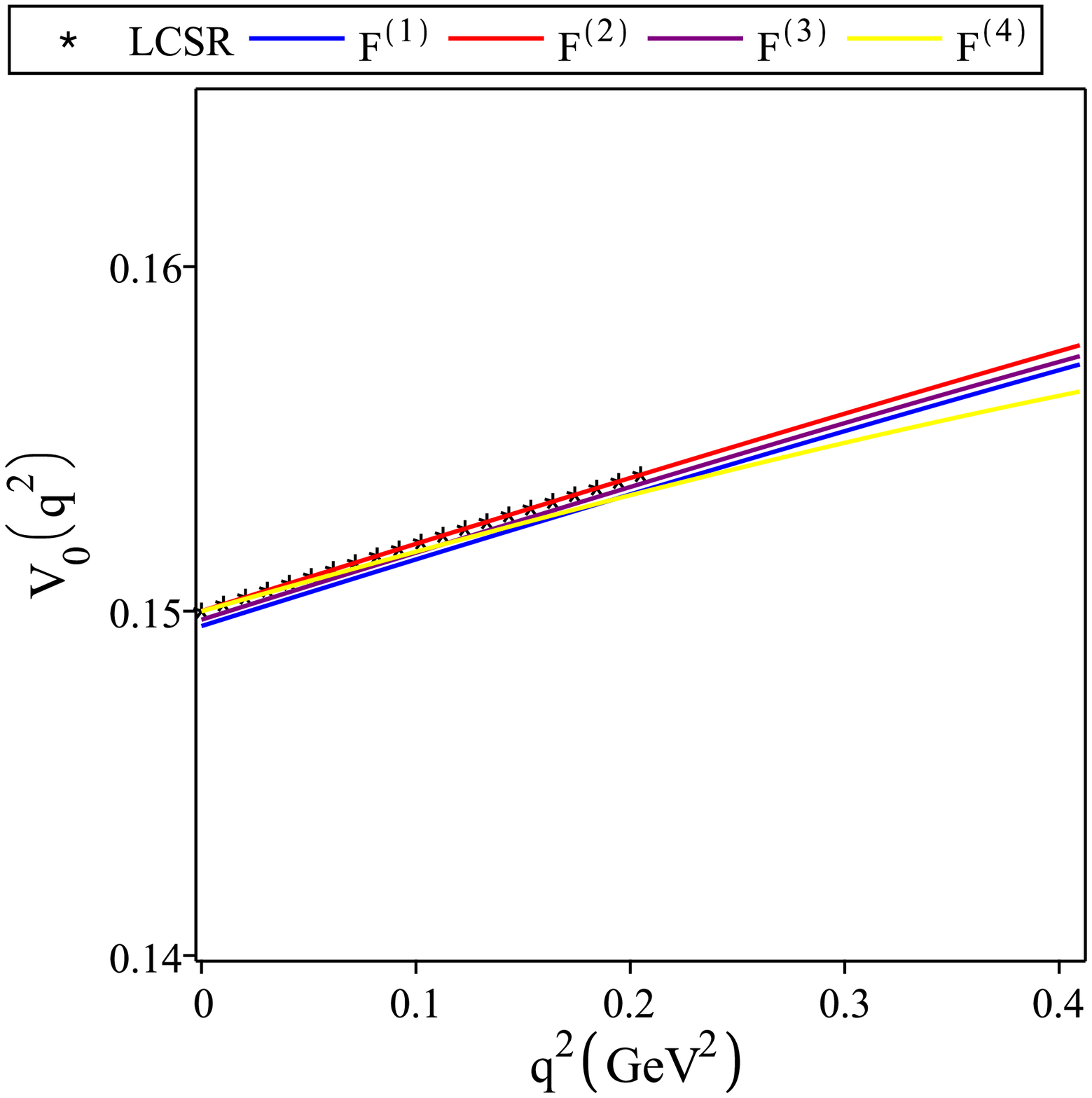}
\caption{Blue, red, purple and yellow lines show the form factors
$A$, $V_{i} (i=0,1,2)$, obtained for $D^0\to a^{-}_{1}$ transition
on $q^2$ by using the  $F^{(1)}$, $F^{(2)}$, $F^{(3)}$ and $F^{(4)}$
fit functions. Asterisks show the results of the LCSR. } \label{ff}
\end{figure}

The dependence of the form factors, $A$, $V_{i} (i=0,1,2)$, for
$D^0\to a^{-}_{1}$ transition on $q^2$ are given in Fig. \ref{ff}.
In this figure the blue, red, purple and yellow lines show the
results for $F^{(1)}$, $F^{(2)}$, $F^{(3)}$ and $F^{(4)}$ fit
functions, respectively. According to the Fig. \ref{ff}, the form
factors obtained for the four fit functions are in a good agreement
with each other and there is no significant change in their
dependence on $q^2$.

The semileptonic $D_{(s)} \to K_{1A} \ell\nu$, $D_{(s)} \to K_{1B}
\ell\nu$, $D^0 \to a^{-}_1 \ell^+ \nu$ and $D^+ \to a^{0}_1 \ell^+
\nu$ decays have been studied via the 3PSR and LFQM approaches
\cite{Khosravi,Chengb,Zuo}. In Tables \ref{T31} and \ref{T32}, we
compare  our results for  the form factors of the aforementioned
decays at the zero transferred momentum square $q^2=0$ with the the
3PSR and LFQM values which have been rescaled according to the form
factor definition in Eq. (\ref{eq27}). In these tables, the errors
of the LCSR values are estimated by the variation of the Borel
parameter $M^{2}$ and variation of the LCDAs parameters. The main
uncertainty comes from the parameters of twist-2 LCDAs.
\begin{table}[th]
\caption{Transition form factors of $D_{(s)} \to K_{1A}(K_{1B})
\ell\nu$ decays at $q^2=0$ in different approaches. } \label{T31}
\begin{ruledtabular}
\begin{tabular}{ccc|cc|cc|cc}
Decay&$D \to K_{1A}$&&$D\to K_{1B}$&&$D_s \to K_{1A}$&&$D_s\to K_{1B}$ \\
\hline
Form factor&This work & 3PSR \cite{Khosravi} &This work & 3PSR \cite{Khosravi}& This work & 3PSR \cite{Khosravi}& This work & 3PSR \cite{Khosravi}\\
\hline
$A(0)$&${0.06}\pm{0.03} $&0.11&${-0.47}\pm{0.14}$ &-0.75&$0.05\pm{0.02}$&0.16&$-0.40\pm{0.11}$&-0.84\\
${V}_{0}(0)$&${0.11}\pm{0.03}$&$0.04$&${-0.42}\pm{0.14}$&$-0.13$&$0.10\pm{0.04}$&0.03&$-0.41\pm{0.12}$&-0.26\\
${V}_{1}(0)$&${0.32}\pm{0.11} $&0.02&${-0.26}\pm{0.10}$&-0.16&$0.28\pm{0.09}$&0.05&$-0.22\pm{0.09}$&-0.30\\
${V}_{2}(0)$&${-0.03}\pm{0.01} $&-0.01&${0.29}\pm{0.13} $&0.08&$-0.01\pm{0.01}$&-0.02&$0.24\pm{0.10}$&0.14\\
\end{tabular}
\end{ruledtabular}
\end{table}
\begin{table}[th]
\caption{Transition form factors of  $D^0\to a^{-}_{1} \ell^+ \nu$
and $D^+\to a^{0}_{1} \ell^+ \nu$ at $q^2=0$ in various theoretical
approaches. } \label{T32}
\begin{ruledtabular}
\begin{tabular}{cccc|ccc}
Decay&$D^0\to a^{-}_{1}$&&&$D^+\to a^{0}_{1}$ \\
\hline
Form factor&This work & 3PSR \cite{Zuo}& LFQM \cite{Chengb}& This work & 3PSR \cite{Zuo}& LFQM \cite{Chengb}\\
\hline
$A(0)$&${0.07}\pm{ 0.05}$&$0.09$&$0.20$&$0.04\pm{0.04}$&0.08&0.14\\
${V}_{0}(0)$&${0.15}\pm{0.05} $&$0.19$&$0.44$&$0.10\pm{0.03}$&0.12&0.30 \\
${V}_{1}(0)$&${0.37}\pm{0.11} $&$0.77$&$1.54$&$0.26\pm{0.08}$&0.54&1.08\\
${V}_{2}(0)$&${-0.03}\pm{0.02}$&$-0.01$&$-0.06$&$-0.02\pm{0.01}$&-0.00&-0.04\\
\end{tabular}
\end{ruledtabular}
\end{table}

The form factor $A(q^2)$ at $q^2=0$ is related to the strong
coupling constant $g_{DD^{*}a_1}$ as
\begin{eqnarray}\label{eq001}
A (0)=\frac{f_{D^*}(m_D-m_{a_1})}{2\,m_{D^*}}g_{DD^{*}a_1}.
\end{eqnarray}
Considering $f_{D^*}=(0.23\pm 0.02)\rm {GeV}$,  the value of
$g_{DD^{*}a_1}$ is evaluated to be $(2.21\pm 1.38)\rm {GeV}^{-1}$.

Now, we study the differential decay widths ${d\Gamma
_{\rm{L}}}/{dq^2}$ and ${d\Gamma _{\pm}}/{dq^2}$ of the semileptonic
decays $D_{(s)}$ to axial vector mesons $A$, given as
\begin{eqnarray}
\frac{d\Gamma_{\rm{L}}(D_{(s)}\rightarrow A \ell \nu)}{dq^2}
&=&{\Big(\frac{q^2-m_l^2}{q^2}\Big)}^2\,\frac{ {\sqrt{\lambda}}\,
G_F^2\, \left| V_{cq'}\right| ^{2}} {384\,\pi^3\,
m_{D_{(s)}}^3}\times \frac{1}{q^2} \Big\{ 3 m_l^2\, \lambda\,
V_0^2(q^2)+ (m_l^2+2q^2) \nonumber\\&& \times \Big|\frac{1}{2m_{A}}
\Big[
(m_{D_{(s)}}^2-m^2_{A}-q^2)(m_{D_{(s)}}-m_{A})V_1(q^2)-\frac{\lambda}{m_{D_{(s)}}-m_{A}}
V_2(q^2)\Big]\Big|^2 \Big\}, \nonumber\\
\frac{d\Gamma_\pm(D_{(s)}\rightarrow A \ell \nu)}{dq^2}
&=&{\Big(\frac{q^2-m_l^2}{q^2}\Big)}^2\,\frac{{\sqrt{\lambda}}
\,G_F^2\,\left| V_{cq'}\right| ^{2}}{384\,\pi^3\, m_{D_{(s)}}^3}
\times \Big\{ (m_l^2+2q^2)\, \lambda
\,\Big|\frac{A(q^2)}{m_{D_{(s)}}-m_{A}}\mp
\frac{(m_{D_{(s)}}-m_{A})V_1(q^2)}{\sqrt{\lambda}}\Big|^2
\Big\},\nonumber
\end{eqnarray}
where $\lambda=m_{D_{(s)}}^4+m_{A}^4+q^4-2\,m_{A}^2\, m_{D_{(s)}}^2
-2\,q^2\, m_{D_{(s)}}^2 -2\,q^2\, m_{A}^2 $. In these relations, for
decays described by $c\to d(s)\, \ell \nu$ transition, $V_{cq'}$
becomes $V_{cd}(V_{cs})$. Also,  ${d\Gamma _{\rm{L}}}/{dq^2}$ and
${d\Gamma _{\pm}}/{dq^2}$ are the longitudinal and transverse
components of the differential decay width, respectively. The total
differential decay width can be written as
\begin{eqnarray}  \label{eq50}
\frac{d\Gamma _{\rm{tot}}(D_{(s)}\rightarrow A \ell
\nu)}{dq^2}=\frac{d\Gamma _{\rm{L} }(D_{(s)}\rightarrow A \ell
\nu)}{dq^2}+\frac{d\Gamma _{\rm{T}}(D_{(s)}\rightarrow A \ell
\nu)}{dq^2},
\end{eqnarray}
where
\begin{eqnarray}  \label{eq51}
\frac{d\Gamma _{\rm{T}}(D_{(s)}\rightarrow A \ell \nu)}{dq^2}
&=&\frac{d\Gamma _{+ }(D_{(s)}\rightarrow A \ell
\nu)}{dq^2}+\frac{d\Gamma _{- }(D_{(s)}\rightarrow A \ell
\nu)}{dq^2}.
\end{eqnarray}

\begin{figure}
\includegraphics[width=6cm,height=6cm]{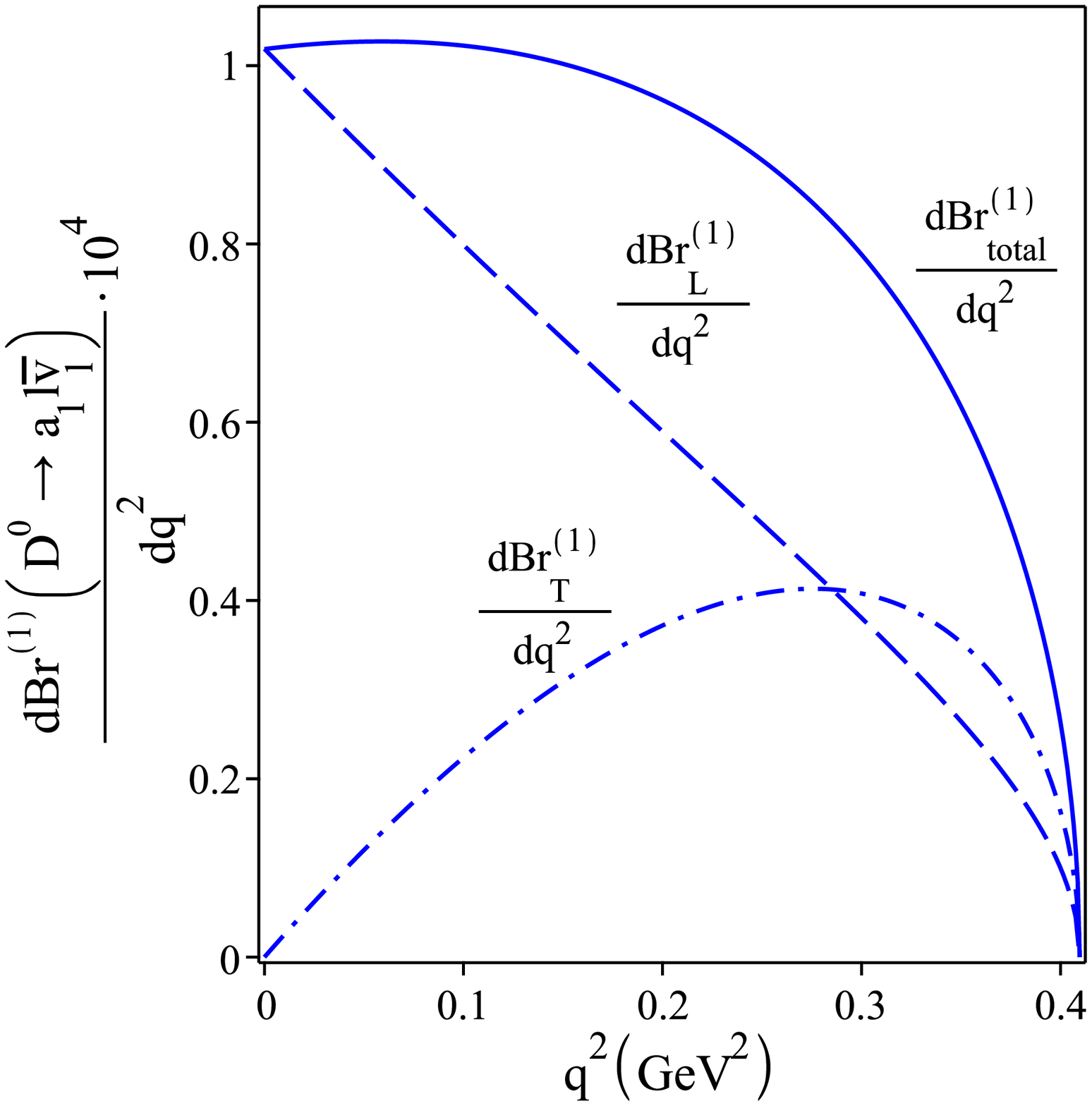}
\includegraphics[width=6cm,height=6cm]{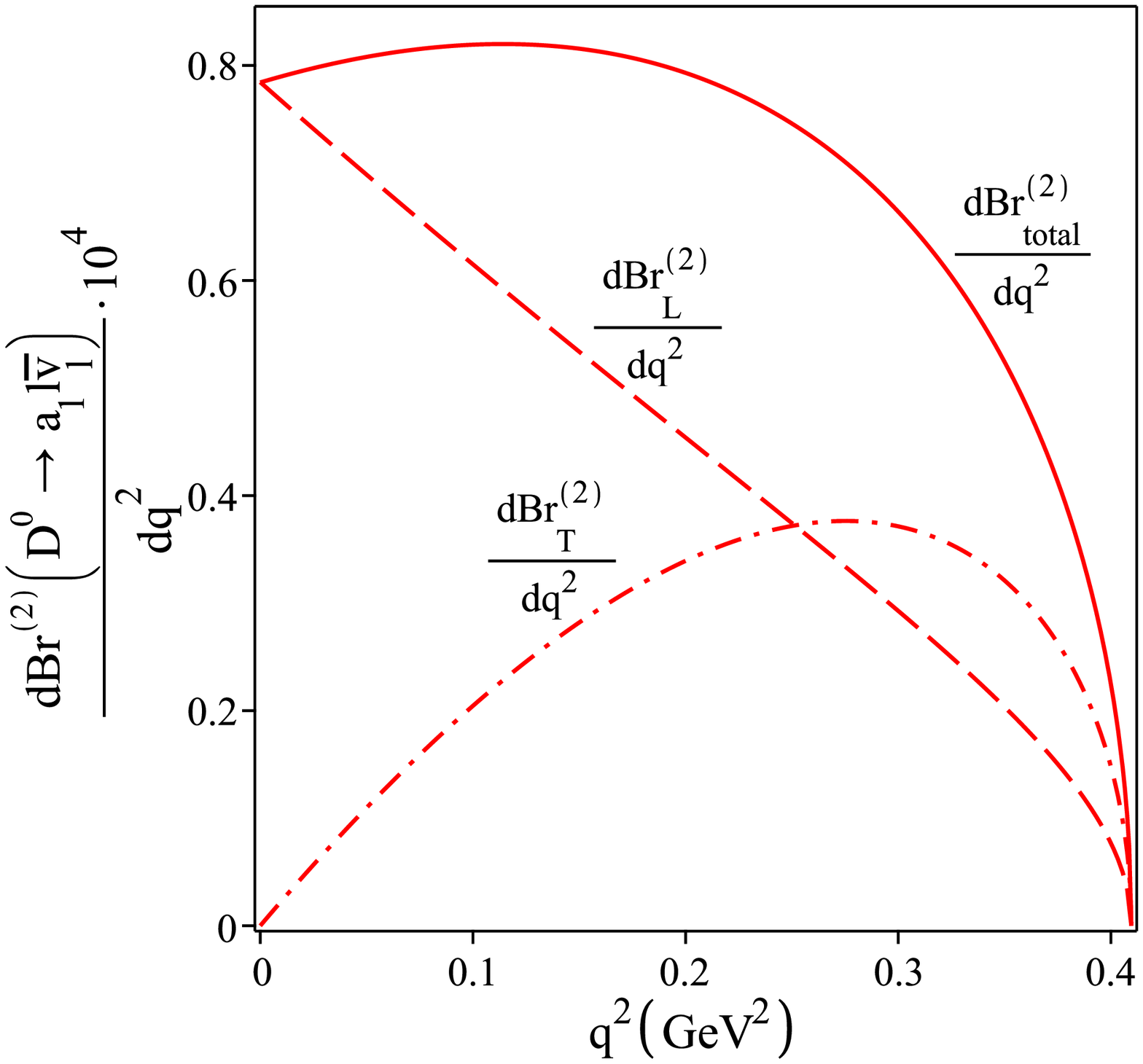}
\includegraphics[width=6cm,height=6.2cm]{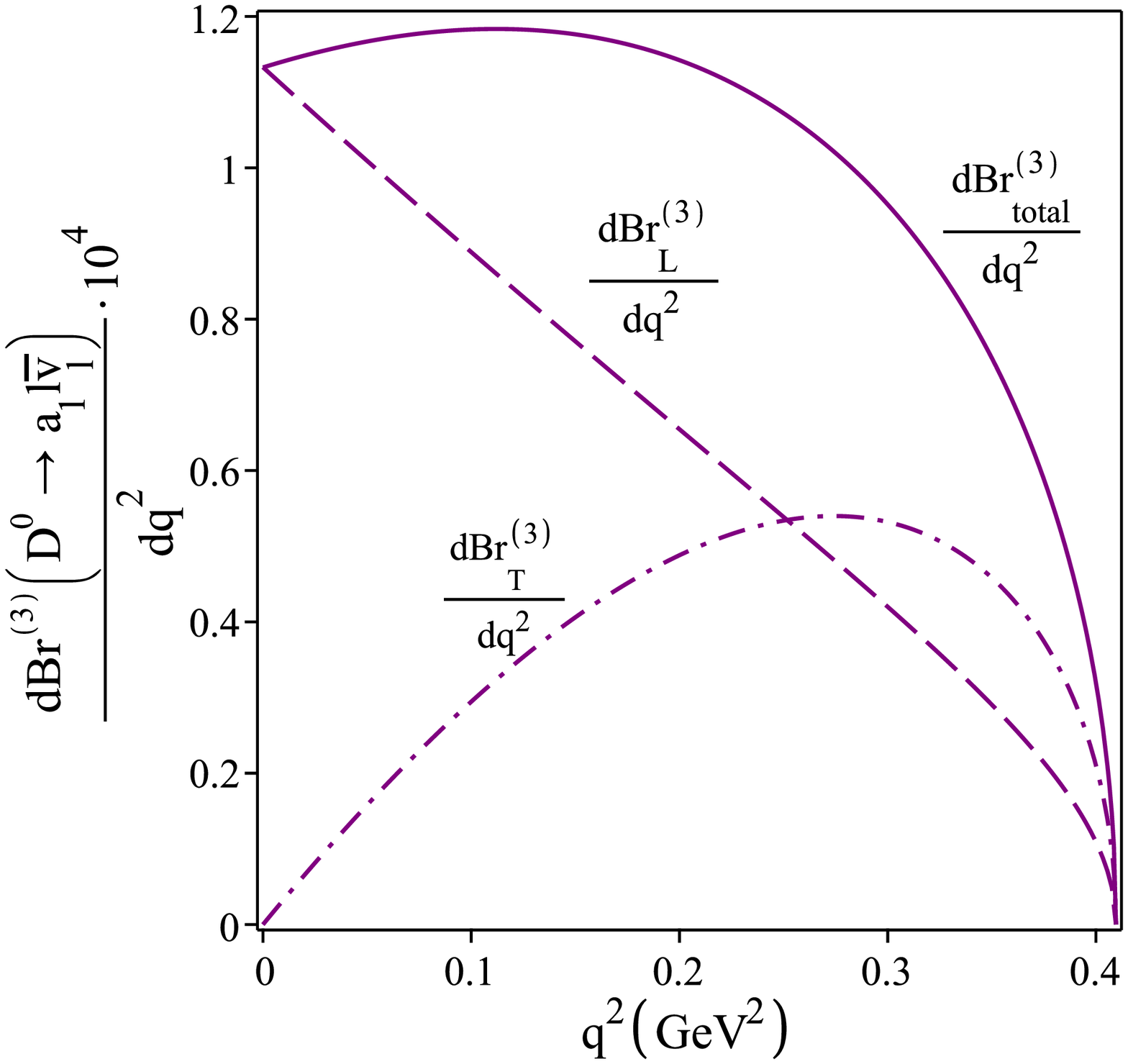}
\includegraphics[width=6cm,height=6.2cm]{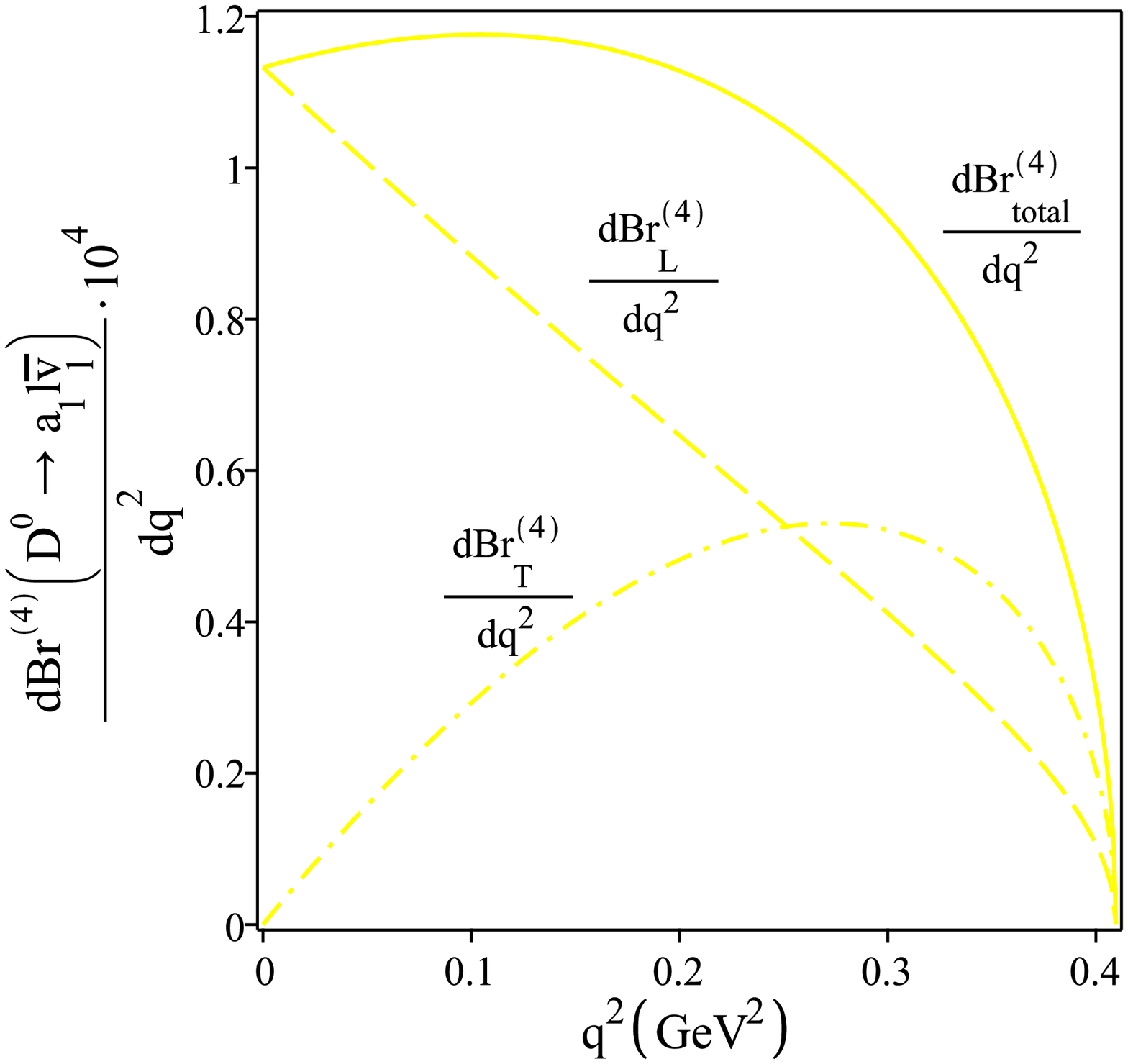}
\caption{The differential branching ratio of $D \to a_{1} \ell \nu$
decay as a function of $q^2$. The solid, dash and dot-dashed lines
depict the total, longitudinal and transverse differential branching
ratio, respectively. Blue, red, purple and yellow plots show the
results using the form factors fitted to $F^{(1)}$, $F^{(2)}$,
$F^{(3)}$ and $F^{(4)}$.} \label{F31}
\end{figure}

We plot the differential branching ratios of  $ D^0 \to a^{-}_{1}
\ell \nu $  with respect to $q^2$ in the physical region
$m_{\ell}^{2}\leq q^2\leq (m_{D^0}-m_{a^{-}})^2$, in Fig. \ref{F31}.
In this figure, the solid, dash and dot-dashed lines depict the
differential branching ratios $dBr_{\rm{tot}}/dq^2$,
$dBr_{\rm{L}}/dq^2$ and $dBr_{\rm{T}}/dq^2$, respectively. Blue,
red, purple and yellow plots show the results of the differential
branching ratios  using the form factors fitted to $F^{(1)},
F^{(2)}, F^{(2)}$ and $F^{(4)}$, respectively.

To calculate the branching ratio values of the semileptonic decays,
we integrate Eq. (\ref{eq50}) over $q^2$ in the whole physical
region and use the total mean life-time $\tau_{D^0}=0.41 $,
$\tau_{D^+}=1.04$ and $\tau_{D_s}=0.50 $ ps \cite{pdg}. To determine
the branching ratio values of $D_{(s)} \to K_1(1270) \ell \nu$ and
$D_{(s)} \to K_1(1400) \ell \nu$ decays, the transition form factors
of them are needed. These form factors can be obtained in terms of
the form factors of  $D_{(s)} \to K_{1A} \ell \nu$ and $D_{(s)} \to
K_{1B} \ell \nu$ decays with the help of the following
transformations:
\begin{eqnarray}
\label{nolabel}
\left( \begin{array}{c}
\langle {K}_1(1270) | \bar{s} \gamma_\mu (1-\gamma_5) c | D \rangle \\
\langle {K}_1(1400) | \bar{s} \gamma_\mu (1-\gamma_5) c | D \rangle
\end{array} \right) =
\left( \begin{array}{lr}
\sin \theta_{K} &  \cos \theta_{K} \\
\cos \theta_{K}   & -\sin \theta_{K}
\end{array} \right)
\left( \begin{array}{c}
\langle {K}_{1A} | \bar{s} \gamma_\mu (1-\gamma_5) c | D \rangle \\
\langle {K}_{1B} | \bar{s} \gamma_\mu (1-\gamma_5) c | D \rangle
\end{array} \right)~.
\end{eqnarray}
As previously mentioned,  we use the result of $\theta_K =
{-(34\pm13)}^{\circ}$.

\begin{figure}
\includegraphics[width=6cm,height=6cm]{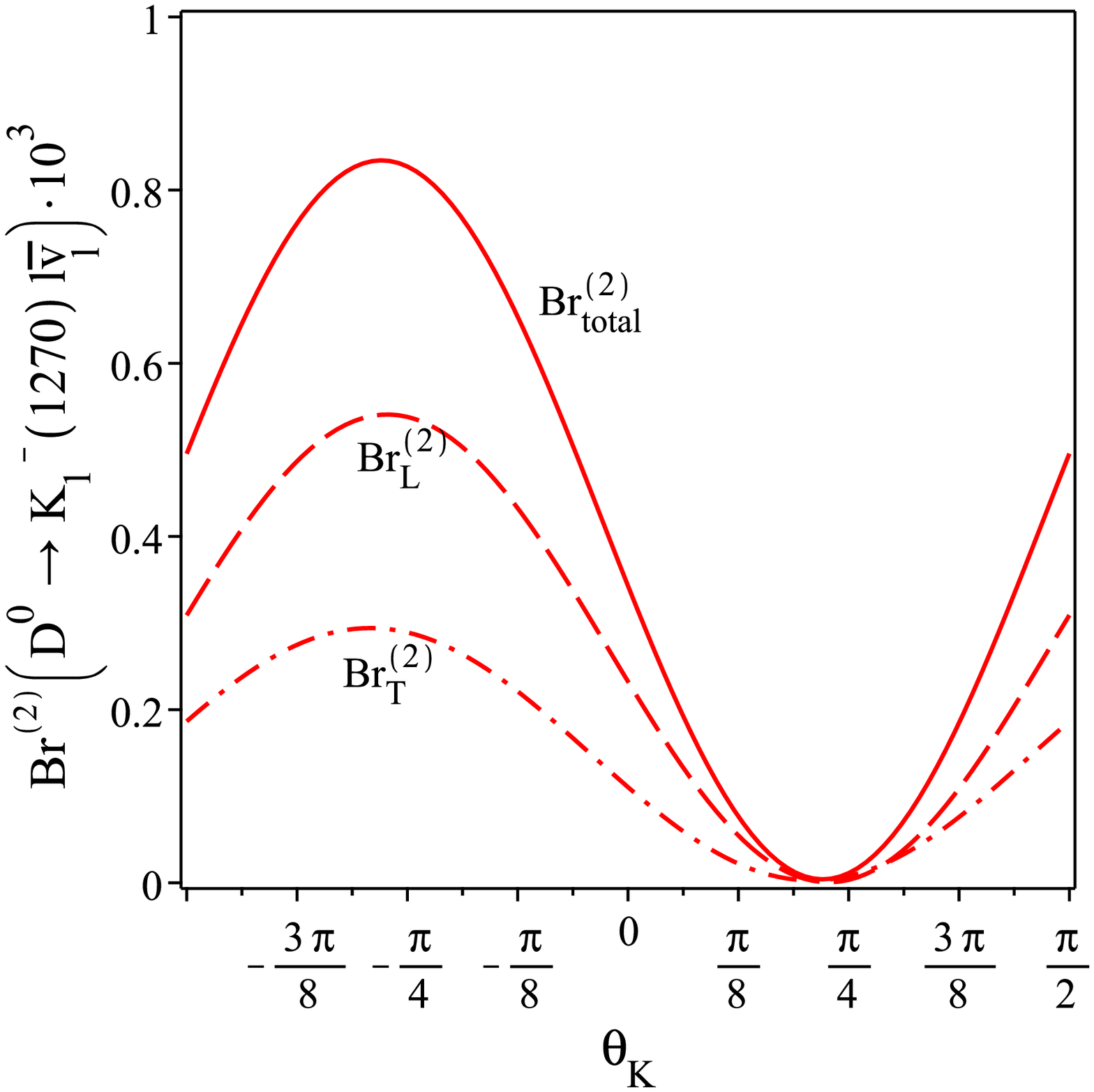}
\includegraphics[width=6cm,height=6cm]{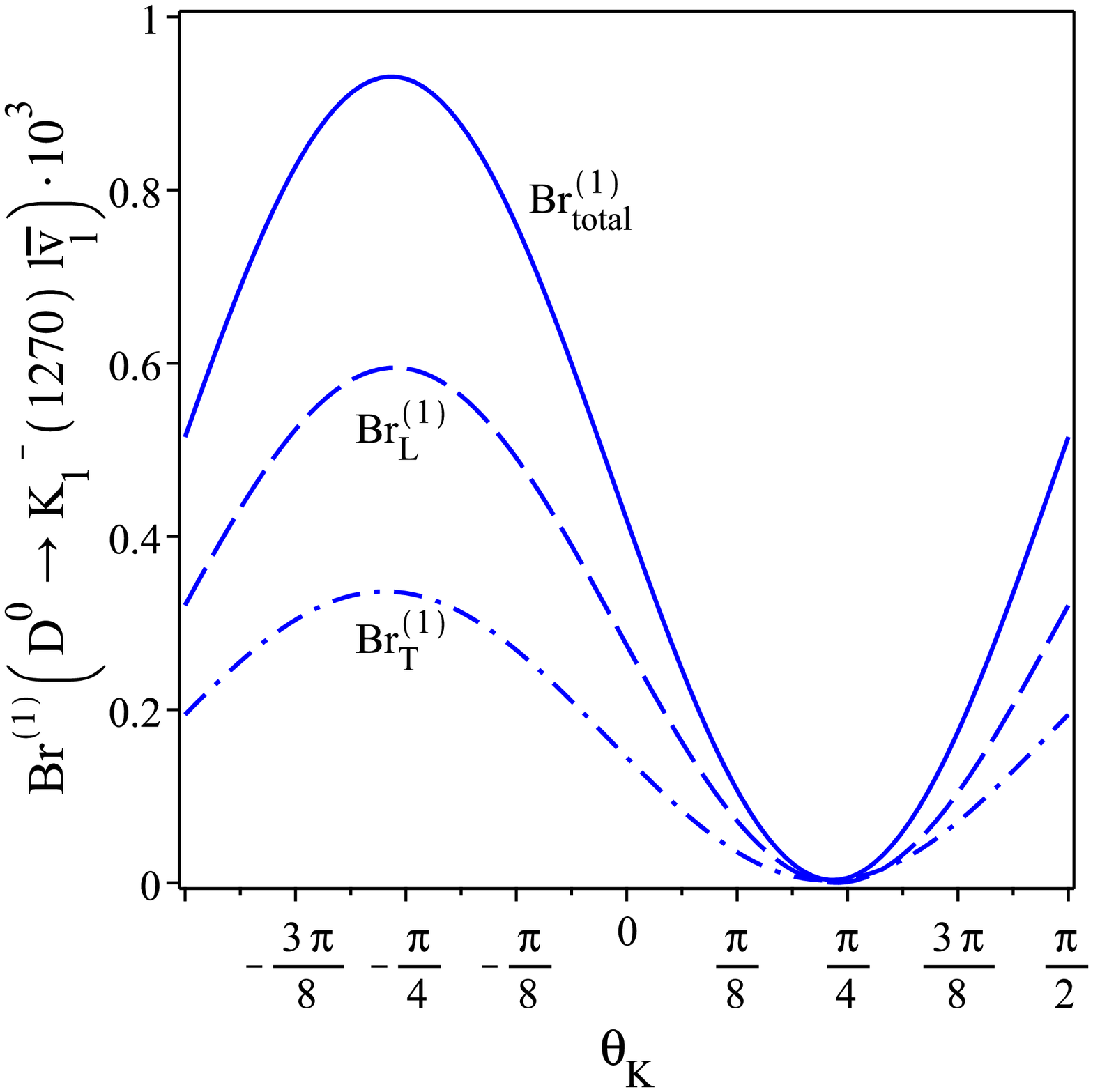}
\includegraphics[width=6cm,height=6cm]{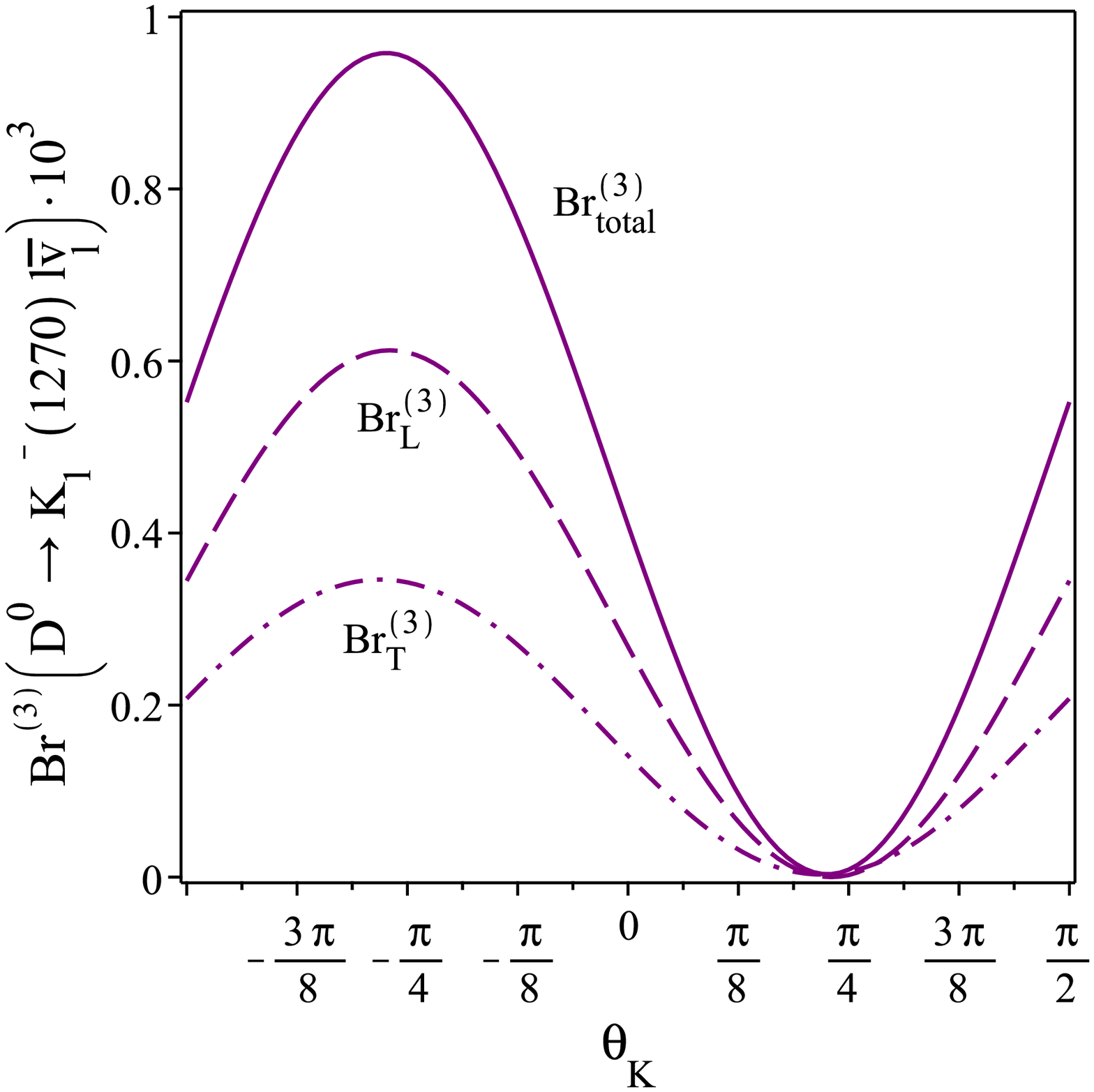}
\includegraphics[width=6cm,height=6cm]{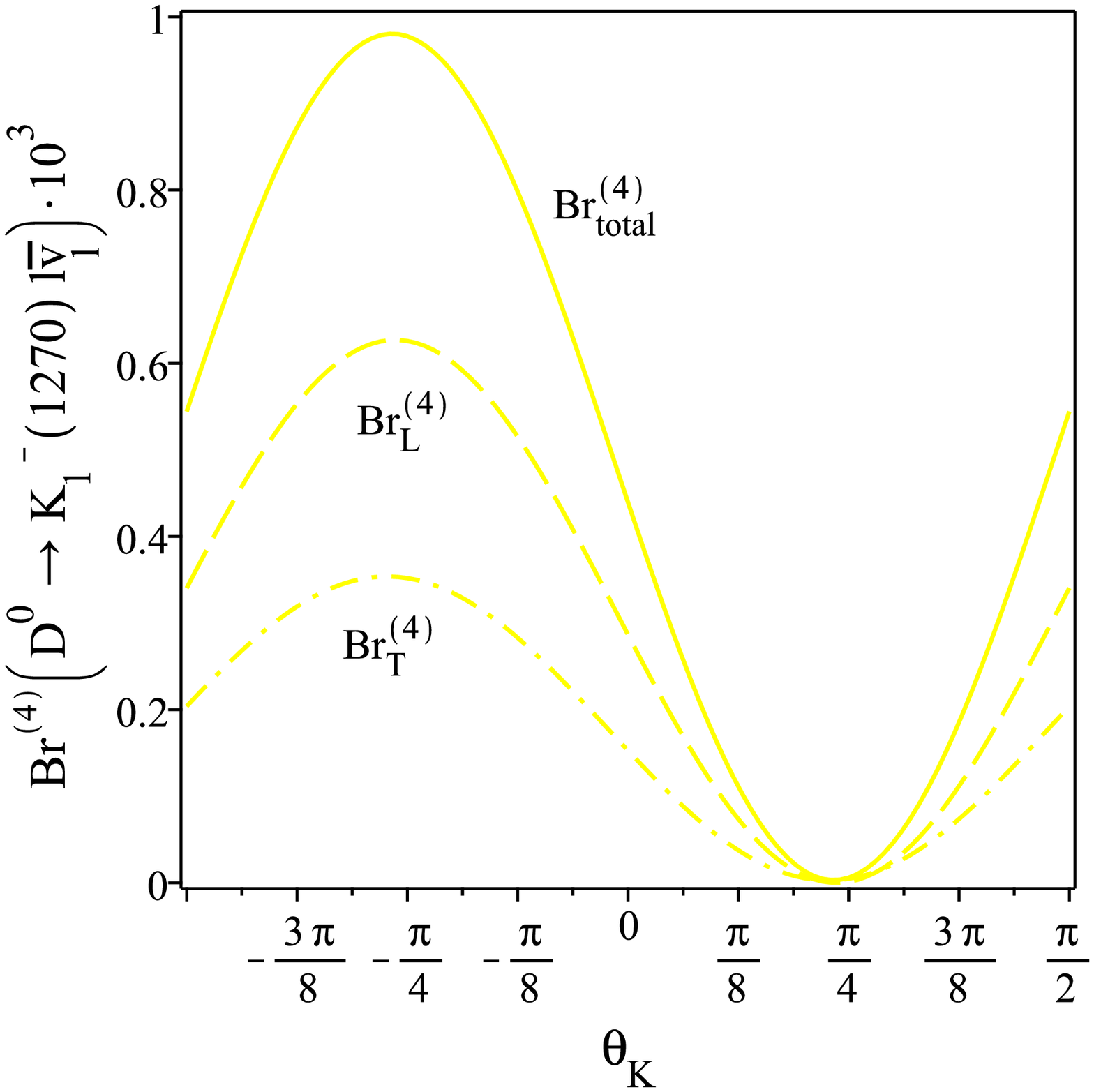}
\caption{The $\theta_{K}$ dependence of branching ratio values of $D
\to K_{1}(1270) \ell \nu$ decay. The solid, dash and dot-dashed
lines depict the total, longitudinal  and transverse branching
ratios, respectively. Blue, red,  purple and yellow plots show the
results using the form factors fitted to $F^{(i)} (i=1,...,4)$.}
\label{F32}
\end{figure}

The branching ratio values of the semileptonic decays $D_{(s)} \to A
\ell \nu$ related to the form factors fitted to $F^{(i)}
(i=1,...,4)$  are presented in Tables \ref{T341}-  \ref{T343},
respectively. In these tables, we compare our results with other
approaches.
\begin{table}[th]
\caption{Branching ratio values of the semileptonic $D_{(s)} \to A
\ell \nu$ related to fit function $F^{(1)}$. } \label{T341}
\begin{ruledtabular}
\begin{tabular}{ccccccccc}
Process& $Br_{_{\rm{T}}}$&$Br_{_{\rm{L}}}$&$Br_{_{\rm{tot}}}(\rm{This~ work})$& $Br_{_{\rm{tot}}}$ \cite{Khosravi} & $Br_{_{\rm{tot}}}$ \cite{Zuo}  \\
\hline
$D^0\to a^{-}_{1}\ell \nu$& $[{1.16} \pm{0.15}$&${2.36} \pm{0.35}$&${3.52} \pm{0.50}$&$--$&$1.11^{+0.41}_{-0.34}$&$]\times10^{-5}$\\
$D^+\to a^{0}_{1}\ell \nu$& $[1.52\pm {0.19}$&$3.05\pm{0.43}$&$4.57\pm{0.62}$&$--$&$1.47^{+0.55}_{-0.44}$&$]\times10^{-5}$\\
$D^0\to b^{-}_{1}\ell \nu$& $[{0.52} \pm{0.08}$&${0.69}\pm {0.10}$&${1.21} \pm{0.18}$&$--$&$--$&$]\times10^{-5}$\\
$D^+\to b^{0}_{1}\ell \nu$& $[0.70\pm{0.11}$&$0.93\pm{0.15}$&$1.63\pm{0.26}$&$--$&$--$&$]\times10^{-5}$\\
$D^0\to K^{-}_{1}(1270)\ell \nu$& $[2.89\pm 0.05$&$5.20\pm 0.10$&$8.09\pm0.15$&$5.34\pm 0.21$&$--$&$]\times10^{-3}$\\
$D^+\to K^{0}_{1}(1270)\ell \nu$& $[7.77\pm 0.12$&$10.82\pm 0.19$&$18.59\pm 0.31$&$14.07\pm 1.22$&$--$&$]\times10^{-3}$\\
$D_{s}^{+}\to K^{0}_{1}(1270)\ell \nu$& $[0.88\pm 0.03$&$1.27\pm 0.06$&$2.15\pm 0.09$&$1.25\pm0.11$&$--$&$]\times10^{-3}$\\
$D^0\to K^{-}_{1}(1400)\ell \nu$& $[0.32\pm 0.03$&$0.43\pm 0.02$&$0.75\pm 0.03$&$0.85\pm 0.02$&$--$&$]\times10^{-3}$\\
$D^+\to K^{0}_{1}(1400)\ell \nu$& $[0.37\pm 0.03$&$0.56\pm 0.03$&$0.93\pm0.06$&$1.27\pm0.10$&$--$&$]\times10^{-3}$\\
$D_{s}^{+}\to K^{0}_{1}(1400)\ell \nu$& $[0.05\pm 0.01$&$0.08\pm
0.01$&$0.13\pm 0.02$&$0.14\pm0.01$&$--$&$]\times10^{-3}$
\end{tabular}
\end{ruledtabular}
\end{table}
\begin{table}[th]
\caption{The same as Table \ref{T341} but related to fit function
$F^{(2)}$. } \label{T34}
\begin{ruledtabular}
\begin{tabular}{ccccccccc}
Process& $Br_{_{\rm{T}}}$&$Br_{_{\rm{L}}}$&$Br_{_{\rm{tot}}}(\rm{This~ work})$& $Br_{_{\rm{tot}}}$ \cite{Khosravi} & $Br_{_{\rm{tot}}}$ \cite{Zuo}  \\
\hline
$D^0\to a^{-}_{1}\ell \nu$& $[{1.06} \pm{0.14}$&${1.79} \pm{0.27}$&${2.85} \pm{0.41}$&$--$&$1.11^{+0.41}_{-0.34}$&$]\times10^{-5}$\\
$D^+\to a^{0}_{1}\ell \nu$& $[1.39\pm {0.18}$&$2.37\pm{0.34}$&$3.76\pm{0.52}$&$--$&$1.47^{+0.55}_{-0.44}$&$]\times10^{-5}$\\
$D^0\to b^{-}_{1}\ell \nu$& $[{0.70} \pm{0.11}$&${1.18}\pm {0.19}$&${1.88} \pm{0.30}$&$--$&$--$&$]\times10^{-5}$\\
$D^+\to b^{0}_{1}\ell \nu$& $[0.91\pm{0.15}$&$1.56\pm{0.25}$&$2.47\pm{0.40}$&$--$&$--$&$]\times10^{-5}$\\
$D^0\to K^{-}_{1}(1270)\ell \nu$& $[2.22\pm 0.03$&$4.56\pm 0.09$&$6.78\pm0.12$&$5.34\pm 0.21$&$--$&$]\times10^{-3}$\\
$D^+\to K^{0}_{1}(1270)\ell \nu$& $[6.56\pm 0.10$&$10.30\pm 0.17$&$16.86\pm 0.27$&$14.07\pm 1.22$&$--$&$]\times10^{-3}$\\
$D_{s}^{+}\to K^{0}_{1}(1270)\ell \nu$& $[0.66\pm 0.02$&$1.00\pm 0.03$&$1.66\pm 0.05$&$1.25\pm0.11$&$--$&$]\times10^{-3}$\\
$D^0\to K^{-}_{1}(1400)\ell \nu$& $[0.34\pm 0.02$&$0.48\pm 0.03$&$0.82\pm 0.05$&$0.85\pm 0.02$&$--$&$]\times10^{-3}$\\
$D^+\to K^{0}_{1}(1400)\ell \nu$& $[0.50\pm 0.04$&$0.78\pm 0.04$&$1.28\pm0.08$&$1.27\pm0.10$&$--$&$]\times10^{-3}$\\
$D_{s}^{+}\to K^{0}_{1}(1400)\ell \nu$& $[0.06\pm 0.01$&$0.10\pm
0.01$&$0.16\pm 0.02$&$0.14\pm0.01$&$--$&$]\times10^{-3}$
\end{tabular}
\end{ruledtabular}
\end{table}
\begin{table}[th]
\caption{The same as Table \ref{T341} but related to fit function
$F^{(3)}$. } \label{T342}
\begin{ruledtabular}
\begin{tabular}{ccccccccc}
Process& $Br_{_{\rm{T}}}$&$Br_{_{\rm{L}}}$&$Br_{_{\rm{tot}}}(\rm{This~ work})$& $Br_{_{\rm{tot}}}$ \cite{Khosravi} & $Br_{_{\rm{tot}}}$ \cite{Zuo}  \\
\hline
$D^0\to a^{-}_{1}\ell \nu$& $[{1.20} \pm{0.16}$&${2.38} \pm{0.36}$&${3.58} \pm{0.52}$&$--$&$1.11^{+0.41}_{-0.34}$&$]\times10^{-5}$\\
$D^+\to a^{0}_{1}\ell \nu$& $[1.57\pm {0.19}$&$3.16\pm{0.44}$&$4.73\pm{0.63}$&$--$&$1.47^{+0.55}_{-0.44}$&$]\times10^{-5}$\\
$D^0\to b^{-}_{1}\ell \nu$& $[{0.67} \pm{0.10}$&${0.80}\pm {0.13}$&${1.47} \pm{0.32}$&$--$&$--$&$]\times10^{-5}$\\
$D^+\to b^{0}_{1}\ell \nu$& $[0.87\pm{0.14}$&$1.04\pm{0.16}$&$1.91\pm{0.30}$&$--$&$--$&$]\times10^{-5}$\\
$D^0\to K^{-}_{1}(1270)\ell \nu$& $[3.18\pm 0.06$&$5.74\pm 0.10$&$8.92\pm0.16$&$5.34\pm 0.21$&$--$&$]\times10^{-3}$\\
$D^+\to K^{0}_{1}(1270)\ell \nu$& $[8.57\pm 0.13$&$11.16\pm 0.19$&$19.73\pm 0.32$&$14.07\pm 1.22$&$--$&$]\times10^{-3}$\\
$D_{s}^{+}\to K^{0}_{1}(1270)\ell \nu$& $[0.96\pm 0.03$&$1.31\pm 0.05$&$2.27\pm 0.08$&$1.25\pm0.11$&$--$&$]\times10^{-3}$\\
$D^0\to K^{-}_{1}(1400)\ell \nu$& $[0.38\pm 0.02$&$0.55\pm 0.05$&$0.93\pm 0.07$&$0.85\pm 0.02$&$--$&$]\times10^{-3}$\\
$D^+\to K^{0}_{1}(1400)\ell \nu$& $[0.56\pm 0.04$&$0.90\pm 0.06$&$1.46\pm0.10$&$1.46\pm0.10$&$--$&$]\times10^{-3}$\\
$D_{s}^{+}\to K^{0}_{1}(1400)\ell \nu$& $[0.08\pm 0.01$&$0.11\pm
0.02$&$0.19\pm 0.03$&$0.14\pm0.01$&$--$&$]\times10^{-3}$
\end{tabular}
\end{ruledtabular}
\end{table}
\begin{table}[th]
\caption{The same as Table \ref{T341} but related to fit function
$F^{(4)}$. } \label{T343}
\begin{ruledtabular}
\begin{tabular}{ccccccccc}
Process& $Br_{_{\rm{T}}}$&$Br_{_{\rm{L}}}$&$Br_{_{\rm{tot}}}(\rm{This~ work})$& $Br_{_{\rm{tot}}}$ \cite{Khosravi} & $Br_{_{\rm{tot}}}$ \cite{Zuo}  \\
\hline
$D^0\to a^{-}_{1}\ell \nu$& $[{1.19} \pm{0.16}$&${2.38} \pm{0.36}$&${3.57} \pm{0.52}$&$--$&$1.11^{+0.41}_{-0.34}$&$]\times10^{-5}$\\
$D^+\to a^{0}_{1}\ell \nu$& $[1.56\pm {0.19}$&$3.14\pm{0.44}$&$4.70\pm{0.63}$&$--$&$1.47^{+0.55}_{-0.44}$&$]\times10^{-5}$\\
$D^0\to b^{-}_{1}\ell \nu$& $[{0.74} \pm{0.12}$&${1.25}\pm {0.20}$&${1.99} \pm{0.32}$&$--$&$--$&$]\times10^{-5}$\\
$D^+\to b^{0}_{1}\ell \nu$& $[0.96\pm{0.16}$&$1.63\pm{0.26}$&$2.59\pm{0.42}$&$--$&$--$&$]\times10^{-5}$\\
$D^0\to K^{-}_{1}(1270)\ell \nu$& $[3.32\pm 0.06$&$5.98\pm 0.11$&$9.30\pm0.17$&$5.34\pm 0.21$&$--$&$]\times10^{-3}$\\
$D^+\to K^{0}_{1}(1270)\ell \nu$& $[8.93\pm 0.14$&$11.68\pm 0.20$&$20.61\pm 0.34$&$14.07\pm 1.22$&$--$&$]\times10^{-3}$\\
$D_{s}^{+}\to K^{0}_{1}(1270)\ell \nu$& $[1.01\pm 0.03$&$1.37\pm 0.06$&$2.38\pm 0.09$&$1.25\pm0.11$&$--$&$]\times10^{-3}$\\
$D^0\to K^{-}_{1}(1400)\ell \nu$& $[0.36\pm 0.02$&$0.52\pm 0.04$&$0.88\pm 0.06$&$0.85\pm 0.02$&$--$&$]\times10^{-3}$\\
$D^+\to K^{0}_{1}(1400)\ell \nu$& $[0.53\pm 0.03$&$0.85\pm 0.05$&$1.38\pm0.08$&$1.27\pm0.10$&$--$&$]\times10^{-3}$\\
$D_{s}^{+}\to K^{0}_{1}(1400)\ell \nu$& $[0.07\pm 0.01$&$0.10\pm
0.01$&$0.17\pm 0.02$&$0.14\pm0.01$&$--$&$]\times10^{-3}$
\end{tabular}
\end{ruledtabular}
\end{table}
The results presented for the branching ratio values of $D\to
K_{1}(1270)\ell \nu$ and $D\to K_{1}(1400)\ell \nu$ decays in Tables
\ref{T341}- \ref{T343} are calculated for $\theta_K =
{-(34\pm13)}^{\circ}$. For a better analysis,   the $\theta_{K}$
dependence of the branching ratio  values of $D\to K_{1}(1270)\ell
\nu$ is displaced in Fig. \ref{F32}.

\subsection{Analysis of nonleptonic decays}

Finally,  we want to evaluate the branching ratio values for the
nonleptonic $D^0 \to K_1^-(1270,1400) \pi^+$ and  $D^+ \to
K_1^0(1270,1400) \pi^+$ decays. The decay width of these nonleptonic
processes is given by:
\begin{eqnarray}
\Gamma(D \to K_1
\pi)=\frac{1}{16\,\pi\,m^{3}_{D}}|\mathcal{M}|^2\,\sqrt{\lambda'}
\end{eqnarray}
where
$\lambda'=m_D^4+m_{K_1}^4+m_\pi^4-2\,m_{K_1}^2\,m^2_D-2\,m_{\pi}^2\,
m^2_D-2\,m_\pi^2\,m_{K_1}^2$. For these decay, the effective
Hamiltonian is given as
\begin{eqnarray}\label{eq3030}
H_{eff} &=&{G_F\over \sqrt{2}}  V^*_{cs} V_{ud}\left(C_1
+\frac{C_2}{N_c}\right)(\bar{s}c)_{V-A}(\bar{u}{d})_{V-A}+h.c.\,.
\end{eqnarray}
In this Hamiltonian, $(\bar{s}c)_{V-A}(\bar{u}{d})_{V-A}=[\bar{s}
\,\gamma^{\mu}(1-\gamma_{5})\,c][\bar{u}\,\gamma^{\mu}(1-\gamma_{5})\,d]$,
and $C_{1}$ and $C_{2}$ are Wilson coefficients. $N_{c}=3$ is the
number of colors in QCD. Using the factorization method, we obtain
the amplitude $\mathcal{M}$  as follows:
\begin{equation}
\mathcal{M}=\sqrt{2}\,G_F  V^*_{cs} V_{ud}\,[e_{1}\,f_{\pi}\,
m_{K_1}\,(\varepsilon^{*}.p_{\pi})\, V_0(m^2_{\pi})],
\end{equation}
where $f_{\pi}$ is the  pion decay constant, and
$e_1=C_1+\frac{1}{N_c} C_2$.  The decay width for $D \to K_1 \pi$
can be written as \cite{Khosravi}:
\begin{eqnarray}\label{eq.gamma1}
\Gamma (D \to K_1 \pi)&=&\frac{G_{F}^{2}}{32~\pi m_{D}^{3}}|V_{cs}|
^{2}|V_{ud }|^{2}~e_{1}^{2}~f_{\pi}^{2}\,\lambda^\frac{3}{2}~
|V_0(m_{\pi}^2)|^{2}.
\end{eqnarray}

To estimate $\Gamma (D \to K_1 \pi)$,  we use $f_{\pi}=0.13
\,\mbox{GeV}$, $m_{\pi}=0.14\,\mbox{GeV}$,  $V_{ud}=0.97$, and
$V_{cs}=0.99$ \cite{pdg}. For obtaining $e_{1}$, the values
$C_1(m_c) = 1.26$ and $C_2(m_c) =-0.51$ are chosen, corresponding to
the results for the Wilson coefficients obtained at the leading
order in renormalization group improved perturbation theory at $\mu
= m_c \simeq 1.4\, \rm {GeV}$ ,in correspondence to $\alpha_s(M_Z) =
0.118$ \cite{Buchalla}. Using four fit functions $F^{(i)}
(i=1,...,4)$, the values for the branching ratios of the nonleptonic
decays $D^0 \to K_1^-(1270) \pi^+$, $D^0 \to K_1^-(1400) \pi^+$,
$D^+ \to K_1^0(1270) \pi^+$ and $D^+ \to K_1^0(1400) \pi^+$ are
obtained and presented in Tables \ref{T01}. This table also contains
the results estimated by the 3PSR method and experiment. As can be
seen in Table \ref{T01}, our results for three fit functions are
close to each other.
\begin{table}[th]
\caption{The branching ratio values of the nonleptonic $D^0 \to
K_1^-(1270) \pi^+$, $D^0 \to K_1^-(1400) \pi^+$, $D^+ \to
K_1^0(1270) \pi^+$ and $D^+ \to K_1^0(1400) \pi^+$ decays via the
different methods and experiment. Our results are related to four
fit functions.} \label{T01}
\begin{ruledtabular}
\begin{tabular}{ccccccc}
Process&  This work ($F^{(1)}$) & This work ($F^{(2)}$)&This work ($F^{(3)}$)&This work ($F^{(4)}$) &3PSR \cite{Khosravi} &  Exp \cite{pdg,Link} \\
\hline
${Br}(D^{0}\to K_{1}^{-}(1270)\pi^{+})\times 10^{-2}$&  $2.55\pm 0.15$   &  $2.57\pm 0.16$   &$2.54\pm 0.15$   &$2.65\pm 0.16$&  $2.26\pm0.18$ & $1.6\pm0.8$\\
${Br}(D^{0}\to K_{1}^{-}(1400)\pi^{+})\times 10^{-2}$&  $0.18\pm 0.03$   &  $0.16\pm 0.03$   &$0.15\pm 0.02$   &$0.17\pm 0.03$&  $0.26\pm0.02$ & $<1.2$       \\
${Br}(D^{+}\to K_{1}^{0}(1270)\pi^{+})\times 10^{-2}$&  $6.19\pm 0.18$   &  $6.42\pm 0.20$   &$6.17\pm 0.20$   &$6.44\pm 0.21$&  $5.85\pm0.37$ & $<0.7$       \\
${Br}(D^{+}\to K_{1}^{0}(1400)\pi^{+})\times 10^{-2}$&  $1.49\pm 0.10$   &  $1.33\pm 0.04$   &$1.30\pm 0.03$   &$1.40\pm 0.08$&  $1.71\pm0.13$ & $3.8\pm 1.3$ \\
\end{tabular}
\end{ruledtabular}
\end{table}

In summary, we investigated the  form factors of the semileptonic
$D_{(s)}$ decay into the $a_1, b_{1}, K_{1}(1270), K_{1}(1400)$
axial vector mesons in the LCSR approach up to the twist--3 LCDAs.
In order to extend our results to the full physical region, we used
four fit functions for parametrization of the form factors. There
was not any significant change in our results using four fit
functions. The branching ratio values of the semileptonic $D^{0} \to
a^{-}_{1} (b^{-}_{1}) \ell^+ \nu$, $D^{+} \to a^{0}_{1} (b^{0}_{1})
\ell^+ \nu$, $D_{s}^{+} \to K^{0}_{1} \ell^+ \nu$ as well as
$D^{+}\to K^{0}_{1} \ell^+ \nu$ decays were evaluated. Using the QCD
factorization method, the nonleptonic decays $D \to K_{1}(1270,1400)
\pi$ were considered and their branching ratio values were
predicted. A comparison was made between our results and other
method predictions and also the experimental values.

\clearpage
\appendix
\section{Twist Function Definitions}\label{app:fun-def}
In this appendix, we present the definitions for the two-- and
three--parton LCDAs as well as the twist functions.

Two--particle chiral--even distribution amplitudes are given by
\cite{Kwei}:
\begin{eqnarray}\label{eq36}
\langle a_1^{-}(p',\varepsilon)|\bar{d}(x) \gamma_\mu \gamma_5
u(0)|0\rangle &=& i f_{a_1^{-}} m_{a_1^{-}}\int_0^1 du \,  e^{i u
p'. x} \Bigg\{ p'_\mu \frac{\varepsilon^{*}. x}{p'. x}
\Phi_\parallel(u) +\left( \varepsilon_{\mu}^{*} -p'_\mu
\frac{\varepsilon^{*}. x}{p'.x}\right) g_\perp^{(a)}(u)
+{\cal O}(x^2) \Bigg\}~,\nonumber\\
\langle a_1^{-} (p',\varepsilon)|\bar{d}(x) \gamma_\mu u(0)|0\rangle
& = & - i f_{a_1^{-}}\, m_{a_1^{-}}\times
\epsilon_{\mu\nu\rho\sigma} \varepsilon^{*\nu} p'^{\rho} x^\sigma
\int_0^1 du \, e^{i u \, p'. x}\Bigg\{
\frac{g_\perp^{(v)}(u)}{4}+{\cal O}(x^2)\Bigg\},
\end{eqnarray}
also, two--particle chiral--odd distribution amplitudes are defined
by:
\begin{eqnarray}\label{eq37}
\langle a_1^{-}(p',\varepsilon)|\bar{d}(x) \sigma_{\mu\nu}\gamma_5
u(0) |0\rangle & =&  f_{a_1^{-}}^{\perp} \int_0^1 du \, e^{i u p'.
x} \Bigg\{(\varepsilon^{*}_{\mu} p'_{\nu} - \varepsilon_{\nu}^{*}
p'_{\mu}) \Phi_\perp(u) + \frac{{m^2_{a_1^{-}}}\,\varepsilon^{*}.
x}{(p'. x)^2}(p'_\mu x_\nu - p'_\nu x_\mu) \bar{h}_\parallel^{(t)}
+{\cal O}(x^2)\Bigg\}, \nonumber \\
\langle a_1^{-}(p',\varepsilon)|\bar{d}(x) \gamma_5 u(0) |0\rangle
&=& f_{a_1^{-}}^\perp m_{a_1^{-}}^2 (\varepsilon^{*}. x)\int_0^1 du
\, e^{i u p'. x}\Bigg\{\frac{h_\parallel^{(p)}(u)}{2}+ {\cal
O}(x^2)\Bigg\}.
\end{eqnarray}
In these expressions, $f_{a_1^{-}}$ and $f_{a_1^{-}}^{\perp}$ are
decay constants of the axial vector meson $a_1^{-}$. We set
$f_{a_1^{-}}^{\perp}=f_{a_1^{-}}$ in $\mu=1~{\rm GeV}$, such that we
have
\begin{eqnarray}
\langle a_1^{-}(p',\varepsilon)| \bar d(0) \sigma_{\mu\nu}\gamma_5
u(0)   |0\rangle  =   a_0^{\perp,a_1^{-}}\,f_{a_1^{-}} \,
(\epsilon^{*}_{\mu} p'_{\nu} - \epsilon_{\nu}^{*} p'_{\mu}),
\end{eqnarray}
where $a_0^{\perp}$ refers to the zeroth Gegenbauer moments of
$\Phi_\perp$. It should be noted that $f_{{a_1^{-}}}$ is
scale--independent and conserves   $G$-parity, but
$f_{{a_1^{-}}}^{\perp}$ is scale--dependent and violates $G$-parity.

We take into account the approximate forms of twist-2 distributions
for the $a_1^{-}$ meson to be \cite{Kwei2}
\begin{eqnarray}\label{eq.t2}
\Phi_\parallel(u) & = & 6 u \bar u \left[ 1 + 3 a_1^\parallel\, \xi +
a_2^\parallel\, \frac{3}{2} ( 5\xi^2  - 1 )
 \right], \\
 \Phi_\perp(u) & = & 6 u \bar u \left[ a_0^\perp + 3 a_1^\perp\, \xi +
a_2^\perp\, \frac{3}{2} ( 5\xi^2  - 1 ) \right],
\end{eqnarray}
where $\xi=2u-1$.

For the relevant two--parton twist--3 chiral--even LCDAs, we take
the approximate expressions up to conformal spin $9/2$ \cite{Kwei2}:
\begin{eqnarray}\label{eq.t3}
g_\perp^{(a)}(u) & = &  \frac{3}{4}(1+\xi^2) + \frac{3}{2}\,
a_1^\parallel\, \xi^3 + \left(\frac{3}{7} \, a_2^\parallel + 5
\zeta_{3,a_1^{-}}^V \right) \left(3\xi^2-1\right)
\nonumber\\
& & {}+ \left( \frac{9}{112}\, a_2^\parallel + \frac{105}{16}\,
\zeta_{3,a_1^{-}}^A - \frac{15}{64}\, \zeta_{3,a_1^{-}}^V
\omega_{a_1^{-}}^V
\right) \left( 35\xi^4 - 30 \xi^2 + 3\right) \nonumber\\
& & + 5\Bigg[ \frac{21}{4}\zeta_{3,a_1^{-}}^V \sigma_{a_1^{-}}^V +
\zeta_{3,a_1^{-}}^A \bigg(\lambda_{a_1^{-}}^A -\frac{3}{16}
\sigma_{a_1^{-}}^A\Bigg) \Bigg]\xi(5\xi^2-3)
\nonumber\\
& & {}-\frac{9}{2} {a}_1^\perp
\,\widetilde{\delta}_+\,\left(\frac{3}{2}+\frac{3}{2}\xi^2+\ln u
+\ln\bar{u}\right) - \frac{9}{2} {a}_1^\perp\,\widetilde{\delta}_-\,
(3\xi + \ln\bar{u} - \ln u),\\
g_\perp^{(v)}(u) & = & 6 u \bar u \Bigg\{ 1 + \Bigg(a_1^\parallel +
\frac{20}{3} \zeta_{3,a_1^{-}}^A
\lambda_{a_1^{-}}^A\Bigg) \xi\nonumber\\
&& + \Bigg[\frac{1}{4}a_2^\parallel + \frac{5}{3}\,
\zeta^V_{3,a_1^{-}} \left(1-\frac{3}{16}\, \omega^V_{a_1^{-}}\right)
+\frac{35}{4} \zeta^A_{3,a_1^{-}}\Bigg] (5\xi^2-1) \nonumber\\
&&+ \frac{35}{4}\Bigg(\zeta_{3,a_1^{-}}^V \sigma_{a_1^{-}}^V
-\frac{1}{28}\zeta_{3,a_1^{-}}^A
\sigma_{a_1^{-}}^A \Bigg) \xi(7\xi^2-3) \Bigg\}\nonumber\\
& & {} -18 \, a_1^\perp\widetilde{\delta}_+ \,  (3u \bar{u} +
\bar{u} \ln \bar{u} + u \ln u ) - 18\, a_1^\perp\widetilde{\delta}_-
\,  (u \bar u\xi + \bar{u} \ln \bar{u} - u \ln u),
\end{eqnarray}
where
\begin{equation}\label{eq.td}
\widetilde{\delta}_\pm  ={f_{a_1^{-}}^{\perp}\over
f_{a_1^{-}}}{m_{u} \pm m_{d} \over m_{a_1^{-}}},\qquad
\zeta_{3,a_1^{-}}^{V(A)} = \frac{f^{V(A)}_{3a_1^{-}}}{f_{a_1^{-}}
m_{a_1^{-}}}.
\end{equation}

Three--particle distribution amplitudes are defined  as:
\begin{eqnarray}\label{eq39}
\langle a_1^{-}(p',\varepsilon) | \bar{d}(x) \gamma_\alpha \gamma_5
g_s G_{\mu\nu} (ux) u (0) | 0 \rangle &=& p'_\alpha (p'_\nu
\varepsilon^{*}_\mu - p'_\mu \varepsilon^{*}_\nu) f_{3a_1^{-}}^A {\cal A} +\cdots,  \nonumber\\
\langle a_1^{-}(p',\varepsilon) | \bar{d}(x) \gamma_\alpha g_s
\widetilde{G}_{\mu\nu} (ux) u (0) | 0 \rangle&= & i p'_\alpha
(p'_\mu \varepsilon^{*}_\nu - p'_\nu \varepsilon^{*}_\mu)
f_{3a_1^{-}}^V {\cal V} +\cdots,
\end{eqnarray}
where
$\widetilde{G}_{\mu\nu}=\frac{1}{2}\epsilon_{\mu\nu\rho\lambda}G^{\rho\lambda}$.

The three--parton chiral--even distribution amplitudes  $\cal A$ and
$\cal V$ in Eq. (\ref{eq39}) are defined as:
\begin{eqnarray}\label{eq310}
{\cal A}&=&  \int {\cal D}\underline{\alpha}
\,e^{ip'.x(\alpha_{1}+u\alpha_{3})}{\cal A} (\alpha_{i}),\nonumber\\
{\cal V}&=&  \int {\cal D}\underline{\alpha}
\,e^{ip'.x(\alpha_{1}+u\alpha_{3})}{\cal V} (\alpha_{i}),
\end{eqnarray}
where ${\cal A}(\alpha_{i})$ and ${\cal V}(\alpha_{i})$ can be
approximately written as \cite{Kwei2}:
\begin{eqnarray}\label{eq311}
{\cal A}(\alpha_{i})&=&5040
(\alpha_{1}-\alpha_{2})\alpha_{1}\alpha_{2}\alpha_{3}^2
+360\,\alpha_{1}\alpha_{2}\alpha_{3}^2 \Big[ \lambda^A_{a_1^{-}}+
\frac{\sigma^A_{a_1^{-}}}{2}(7\alpha_3-3)\Big],
\nonumber\\
{\cal V}(\alpha_{i})&=&360\,\alpha_{1}\alpha_{2}\alpha_{3}^2 \Big[
1+ \frac{\omega^V_{a_1^{-}}}{2}(7\alpha_3-3)\Big] +5040
(\alpha_{1}-\alpha_{2})\alpha_{1}\alpha_{2}\alpha_{3}^2
\sigma^V_{a_1^{-}},
\end{eqnarray}
In these expressions $\alpha_{1}$, $\alpha_{2}$, and $\alpha_3$ are
the momentum fractions carried by $d$, $\bar u$ quarks and gluon,
respectively, in the axial vector meson $a_1^{-}$. The integration
measure is defined as:
\begin{equation}\label{eq312}
\int {\cal D}\underline{\alpha} \equiv \int_0^1 d\alpha_{1} \int_0^1
d\alpha_{2}\int_0^1 d\alpha_3 \,\delta(1-\sum \alpha_i).
\end{equation}

\clearpage
\section{Form Factor Expressions }\label{app:form factors}

In this appendix, the explicit expressions for the form factors of
the semileptonic $D^0 \to a_1^{-} \ell^{+} \nu$ decay are presented.
\begin{eqnarray*}
V_{1}(q^{2})&=&
-\frac{m_{c}\,f_{a_1^{-}}^{\perp}}{8\,m_{D^0}^2(m_{D^0}-m_{a_1^{-}})\,f_{D^0}}\,e^{\frac{m_{D^0}^2}{M^2}}\Bigg\{\hat{\mathcal{L}}\Bigg[7
~\frac{\Phi_\perp(u)\,\delta_{1}(u)}{2\,u}+2\,m_{a_1^{-}}^{2}\frac{h_\parallel^{(p)}(u)}{u}
-3\,\frac{m_{c}\,m_{a_1^{-}}\,f_{a_1^{-}}}{f_{a_1^{-}}^{\perp}}\frac{{g_\perp^{(a)}(u)}}{u}\nonumber\\
&-&m_{a_1^{-}}^{2} \frac{\bar{h}{_\parallel^{(t)(ii)}}(u)}{u^{2}}
\,(7+\frac{\delta_{2}(u)}{M^2})-
4\,\frac{m_{c}\,m_{a_1^{-}}^{3}}{M^2}\frac{\phi_{b}^{ii}(u)}{u^2}\Bigg]e^{s(u)}
+4\,\frac{m_{c}\,m_{a_1^{-}}^{2}}{M^{2}\,f_{a_1^{-}}^\perp
}\hat{\mathcal{L}}\Bigg[\int {\cal
D}\,\underline{\alpha}\left(\frac{\delta_{3}(\alpha_{i})}{\kappa^{2}}\right)e^{s(\kappa)}\Bigg]\Bigg\},\nonumber \\
V_{2}(q^{2})&=&
\frac{2\,m_{c}(m_{D^0}-m_{a_1^{-}})\,f_{a_1^{-}}^{\perp}}{m_{D^0}^2\,f_{D^0}}\,e^{\frac{m_{D^0}^2}{M^2}}
\,\Bigg\{\hat{\mathcal{L}}\Bigg[8~ \frac{\Phi_\perp (u)}{u}+2\,
\frac{m_{c}\,m_{a_1^{-}}\,f_{a_1^{-}}}{M^2\,f_{a_1^{-}}^{\perp}}\frac{\phi_{a}(u)}{u^2}
+2\,m_{a_1^{-}}^{2}\,(1+2u)\,\frac{{h_\parallel^{(p)}(u)}}{u} \nonumber\\
&+&8\,\frac{m_{c}\,m_{a_1^{-}}\,f_{a_1^{-}}}{
M^{2}\,f_{a_1^{-}}^{\perp}} \frac{{\Phi_{\|}}^{(i)}(u)}{u^{2}}
-2\,\frac{m_{a_1^{-}}^{2}}{M^2}
\frac{\bar{h}{_\parallel^{(t)(ii)}}(u)}{u^{3}}\left(\frac{\delta_{4}(u)}
{M^2}-7u-2\right)-8\,\frac{m_{c}\,m_{a_1^{-}}^{3}}{M^4}\frac{\phi_{b}^{ii}(u)}{u^3}\Bigg] e^{s(u)}\Bigg\}\nonumber\\
V_{0}(q^{2})-V_{3}(q^{2})&=&q^2\,\frac{m_{c}\,f_{a_1^{-}}^{\perp}}{8\,m_{D}^2\,m_{a_1^{-}}\,f_{D}}
\,e^{\frac{m_D^2}{M^2}}\,\Bigg\{\hat{\mathcal{L}}\Bigg[8\,\frac{\Phi_\perp
(u)}{u}+2\,\frac{m_{c}\,m_{a_1^{-}}\,f_{a_1^{-}}}{M^2\,f_{a_1^{-}}^{\perp}}
~\frac{\phi_{a}(u)}{u^2}-4\, m_{a_1^{-}}^{2}
\frac{{h_\parallel^{(p)}(u)}}{u}~(1-u)
\nonumber\\
&&+16\,\frac{m_{c}\,m_{a_1^{-}}\,f_{a_1^{-}}}{M^{2}\,f_{a_1^{-}}^{\perp}}
\frac{{\Phi_{\|}^{(i)}}(u)}{u^{2}} +2\, m_{a_1^{-}}^{2}\,
\frac{\bar{h}_{\parallel}^{(t)(ii)}(u)}{u^{3}}\left(\frac{\delta_{4}(u)-M^2(7u+2)}{M^4}
\right)-4\,\frac{m_{c}\,m_{a_1^{-}}^{3}\,f_{a_1^{-}}}{M^4\,f_{a_1^{-}}^{\perp}}\nonumber\\
&\times&\frac{\phi_{b}^{ii}(u)(1-u)}{u^4}\Bigg]e^{s(u)}\Bigg\},
\end{eqnarray*}
where
\begin{eqnarray*}
\hat{\mathcal{L}}&=&\int_{u_0}^{1}du\,,\\
u_{0} &=&\frac{1}{2m_{a_1^{-}}^2} \left[\sqrt{(s_0-m_{a_1^{-}}^2-q^2)^2 +4 m_{a_1^{-}}^2 (m_c^2-q^2)} -\left(s_0-m_{a_1^{-}}^2-q^2\right)\right],\nonumber\\
s(u)&=&-\frac{1}{M^2\,u}\left[m_c^2+u\,\bar{u}m_{a_1^{-}}^2-\bar{u}q^2\right],\nonumber\\
\delta_{1}(u)&=& m_{a_1^{-}}^2(u+2)+\frac{m_{c}^2}{u}+\frac{q^2}{u}\nonumber,\nonumber\\
\delta_{2}(u)&=& m_{a_1^{-}}^2\,u+\frac{m_{c}^2}{u}-\frac{q^2}{u}\nonumber,\nonumber\\
\delta_{3}(\alpha_{i})&=&f_{3a_1^{-}}^A\,{\cal A}(\alpha_{i})-
f_{3a_1^{-}}^V\,{\cal V}(\alpha_{i}),\nonumber\\
\delta_{4}(u)&=&\frac{m_{c}^2}{u}(2-16u)+2m_{a_1^{-}}^2(2+u(1-u))-\frac{q^2}{u}(15+12u),\nonumber\\
{h}^{(i)}(u)&\equiv&\int_0^u h(v) dv,\nonumber\\
{h}^{(ii)}(u)&\equiv&\int_0^u dv\int_0^v d\omega~ h(\omega),\nonumber\\
\phi_a&=& \int_0^u \left[\Phi_\parallel - g_\perp^{(a)} (v)\right]dv,\nonumber\\
\kappa&=&\alpha_1+u\alpha_3.
\end{eqnarray*}

\end{document}